\documentclass[12pt]{article}
\usepackage{preamble_PLM}

\title{Inference in Partially Linear Models under Dependent Data with Deep Neural Networks}
\date{ 
    \vspace{-.2cm} 
    \today 
    \vspace{-.8cm}
}

\author{
Chad Brown%
    \thanks{Department of Economics, University of Colorado Boulder, 80309, USA. 
    chad.brown@colorado.edu.} 
}

\begin{document} 
    \setlength{\abovedisplayskip}{4pt}
    \setlength{\belowdisplayskip}{4pt}
\maketitle

\begin{abstract}
\vspace{-.3cm}
\renewcommand{\thefootnote}{\fnsymbol{footnote}}
\singlespacing
    \noindent 
    I consider inference in a partially linear regression model under stationary $\beta$-mixing data after first stage deep neural network (DNN) estimation.
    Using the DNN results of \citet{brown_statistical_2024}, I show that the estimator for the finite dimensional parameter, constructed using DNN-estimated nuisance components, achieves $\sqrt{n}$-consistency and asymptotic normality.
    By avoiding sample splitting, I address one of the key challenges in applying machine learning techniques to econometric models with dependent data. 
    In a future version of this work, I plan to extend these results to obtain general conditions for semiparametric inference after DNN estimation of nuisance components, which will allow for considerations such as more efficient estimation procedures, and instrumental variable settings.
    \\
    
    
\end{abstract}

\newpage

\section{Introduction}\label{sec:intro}
    Partially linear models were first considered by 
        \citet{engle_semiparametric_1986}
    and
        \citet{robinson_1988}.
    Since then, these models have
    been widely used for empirical work in time series settings (see e.g., \citealp{gao_nonlinear_2007}; and \citealp{hardle_partially_2000}; for a review). 
    For instance, partially linear models have been employed by
    \citet{engle_semiparametric_1986} 
    to study electricity sales,
        since the impact of temperature on electricity consumption is nonlinear, as both high and low temperatures lead to increased electricity demand;
    \citet{li_simultaneous_2024} 
        to study the forward premium anomaly;
    and
    \citet{gao_adaptive_2000}
        to
        study the number of lynx trapped in the MacKenzie River district in the Canadian North-West Territories.

    In particular, consider the partially linear model,
        \begin{equation}\label{eq:PLR1}
        \begin{aligned}
            \Yt 
        = 
            \Tt\paramO + \nus(\Xt) + u_t,
        \qquad
        \text{ where }
        \quad
            \E[u_t | \Tt,\Xt]= 0
        .
        \end{aligned}
        \end{equation}
    Let $\{\Zt\coloneqq(\Yt,\Tt,\Xt)\}_{t\in\N}$ be an stochastic sequence on the probability space
    $\probspace$.
    The data is 
        $\{\Zt\}_{t=1}^n$,
    where 
        $\Yt$ 
    is the outcome,
        $\Tt$
    is the policy or treatment variable, and
        $\Xt$
    are additional covariates 
    which affect $\Yt$ through an unknown nuisance function 
    $\nus$ that is measurable.
    Note that this model can apply to nonlinear auto-regressive settings by letting $\Xt$ include past values $\Y_{t-j}$, $j=1,\ldots,r$ for $r\in\N$.
    For notational simplicity, we consider the case where $\Tt$ is a scalar treatment, although the results obtained here could easily generalize to the vector-valued case. 
    We are interested in estimating and performing inference on the parameter $\paramO\in\R$ in settings where 
    $\{\Zt\}_{t\in\N}$ is (strictly) stationary and $\beta$-mixing (see e.g. \citealp[Definition 3.1, p.19]{dehling_empirical_2002}), 
    in the sense that $\beta(j)\to0$ as $n\to\infty$ for
        $$
                \beta(j) 
            \coloneqq
                \E\Bigg[
                    \sup\bigg\{
                    \Big|
                        \P\big(
                            B|\,
                            \sigma\big(\{\Zt\}_{1}^k\big)
                        \big)
                        -
                        \P\big(
                            B
                        \big)
                    \Big|
                    :\,
                    {B\in \sigma\left(\{\Zt\}_{k+j}^\infty\right)}
                    ,\,
                    k\in\N
                    \bigg\}
                \Bigg]
            ,
            \quad
            \text{ for }
            j\in\N
            ,
            $$
    where 
        $\sigma\left(\{\Zt\}_{k}^n\right)$ 
    denotes the $\sigma$-algebra generated by $\{\Zt\}_{t=k}^n$.

    Using a procedure due to \citet{robinson_1988}, 
    the estimator for $\paramO$ will be constructed using DNN-estimated nuisance components. 
    With the DNN results from \citet{brown_statistical_2024}, I show that this estimator will obtain $\sqrt{n}$-consistency and asymptotic normality in settings with stationary $\beta$-mixing data under mild regularity conditions.
    By employing the ideas of \citet[Theorem 1]{chen_debiased_2022}, I do this
    without sample splitting, which is particularly important in dependent data settings, where it can be difficult to construct independent validation sets.

    This work primarily contributes to two growing bodies of literature.
    First, the literature on inference after machine learning (e.g.,  \citealp{chernozhukov_doubledebiased_2018}; \citealp{farrell_deep_2021};
    \citealp{chernozhukov_automatic_2022})
    largely focuses on \iid settings, whereas I provide results that deliver valid large sample inference under dependent data settings.
    Second,
    the
    work that studies partially linear models 
    in \iid settings 
        (see e.g., \citealp{robinson_1988}; 
        \citealp{fan_profile_2005};
        \citealp{geng_estimation_2020}; and citations therein)
    and time series settings
        (e.g., 
        \citealp{hardle_partially_2000};
        \citealp{gao_nonlinear_2007}; 
        \citealp{li_simultaneous_2024})
    has used more traditional nonparametric estimators, 
    such as kernel estimators,
    whereas I use DNNs to estimate the infinite dimensional parameters.
    In addition, it also shows practical implications of the results from \citet{brown_statistical_2024}, and introduces techniques that could be applied in settings beyond the simple partially linear model considered here.

    {
\renewcommand{\Xspace}{\mathbb{X}}
    \textbf{Notation:}
    For two sequences of non-negative real numbers 
        $\{x_t\}_{t\in\mathbb{N}}$ and $\{y_t\}_{t\in\mathbb{N}}$, 
    the notation 
        $x_t\lesssim y_t$ 
    means there exists a constant $0<C<\infty$ such that    
        $x_t\leq Cy_t$
    for all $t$ sufficiently large. 
    We write 
        $x_t \asymp y_t$
    if 
        $x_t \lesssim y_t$ and $x_t \gtrsim y_t$.
    For some set $\Xspace$, and $A\subseteq\Xspace$, the indicator function is denoted as
            $\mathbbm{1}_{A}:\Xspace\to\{0,1\}$,
        where 
            $\mathbbm{1}_{A}(x)=1$ if $x\in A$ 
        and
            $\mathbbm{1}_{A}(x)=0$ if $x\in \Xspace\setminus A$.
    For a measurable space $(\Xspace, \Sigma)$,
        and a random variable
            $\X:\Omega \to \Xspace$,
        define the measure 
            $\Px(B) = \P(\X^{-1}(B))$
        for any $B\in \Sigma$.
    Let 
                $\Lp{r}\probspace$
            denote the space of functions 
                $
                    f:
                    \Omega
                    \to
                    \R
                $
            that are 
            Borel measurable and
                $\norm{f}_{\Lp{r}\probspace}<\infty$,
            for the (pseudo-) norms
                \begin{equation*}
                    \norm{f}_{\Lp{r}\probspace} 
                \coloneqq
                    \left\{
                    \begin{aligned}
                        &
                        \left(\int_{\Omega}|f|^rd\P\right)^{1/r},
                        \qquad &\text{for } 1 \leq r < \infty,
                        \\&
                        \inf\Big\{C\geq0: 
                        \P(\{\omega\in\Omega:|f(\omega)|\geq C\}) = 0
                        \Big\},
                        &\text{for } r = \infty.
                    \end{aligned}
                    \right.
                \end{equation*}
            We write 
                $\Lp{r}(\P)$ or $\norm{f}_{\Lp{r}(\P)}$
            when no confusion may arise. 
            
}

{
\section{Inference in Partially Linear Models}

    We estimate $\paramO$ following a procedure due to \citet{robinson_1988}.
    Note that  
        $$
        \E[\Yt|\Xt] = \E[\Tt|\Xt]\paramO + \nus(\Xt) + \E[u_t|\Xt]
        .
        $$
    Let 
        $\fOY\coloneqq\E[\Yt|\Xt]$ and
        $\fOT\coloneqq\E[\Tt|\Xt]$,
    then, taking the difference of \eqref{eq:PLR1} and the previous display,
        \begin{equation*}
        \begin{aligned}
            \Yt - \fOY(\Xt)
        = 
            \big(\Tt-\fOT(\Xt) \big)\paramO + \upsilon_t,
        \qquad
        \text{ where }
        \quad
            \E[\upsilon_t | \Tt,\Xt]= 0.
        \end{aligned}
        \end{equation*}
    This provides the moment condition 
        $
            \E\big[\mom{\Zt}{\paramO}{\fOT}{\fOY}\big]
        =
            0
        ,
        $
    for the linear moment function
        \begin{equation*}
        \begin{aligned}
            \mom{\Zt}{\param}{\ftnT}{\ftnY}
        &
        \coloneqq
            \big(\Tt-\ftnT(\Xt)\big)
            \Big[
            \big(\Tt-\ftnT(\Xt)\big)\param
            -
            \Yt - \ftnY(\Xt)
            \Big]
        \\&
        =
            \big[
                \Tt - \ftnT(\Xt)
            \big]^2
            \param
            -
            \big[
                \Tt - \ftnT(\Xt)
            \big]
            \big[
                \Yt - \ftnY(\Xt)
            \big]
        .
        \end{aligned}
        \end{equation*}
    Now, we construct a two-stage method of moments estimator for $\paramO$. 
    In the first stage,
    we estimate the conditional expectations $\fOY$ and $\fOT$ with DNN estimators $\fY$ and $\fT$ constructed using the framework of Section \ref{sec:DNN}.
    Then, we have the estimator
        \begin{equation*}
            \paramhat
        =
            \left( 
                \sumin
                \big[
                    \Tt - \fT(\Xt)
                \big]^{2}
            \right)^{-1}
            \left( 
                \sumin
                \big[
                    \Tt - \fT(\Xt)
                \big]
                \big[
                    \Yt - \fY(\Xt)
                \big]
            \right)
        ,
        \end{equation*}
    which exists whenver 
        $\big[\Tt-\fT(\Xt)\big]^2\neq 0$ 
    for atleast one $t\in\{1,\ldots,n\}$.

\subsection{First Stage DNN Estimation}\label{sec:DNN}
    This section considers the preliminary DNN estimation of the conditional expectation functions $\fOT$ and $\fOY$.

    \subsubsection{DNN Architectures}\label{sec:DNN_arch}
\begin{figure}[t]
    \centering
    \includegraphics[width =0.5\textwidth, height = 4cm]{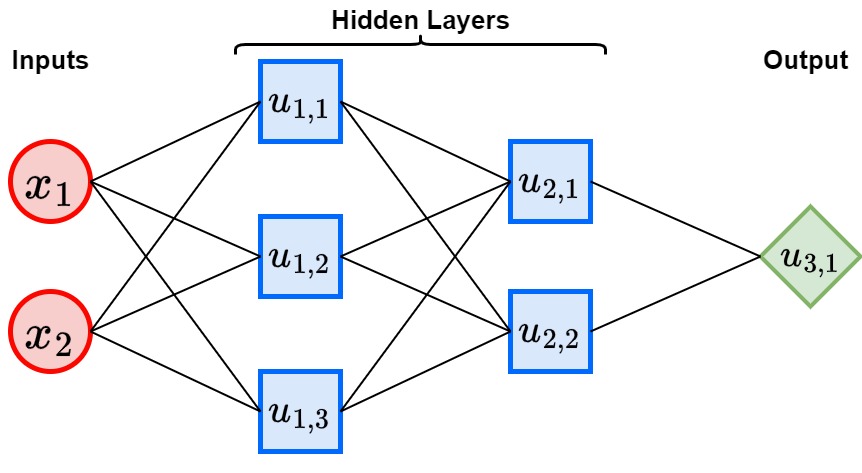}
    \caption{Example of $\FMLP$ architecture graph structure where 
        $\L=2$, $\Hi{1} =3$, $\Hi{2}=2$, $\W=20$, and $d=2$.}
    \label{fig:DNN_MLP}
\end{figure}
    First, we describe the DNN architectures used for the first stage DNN estimators.
    This will follow the framework of 
    \citet[\S3.1]{brown_statistical_2024}. 
    The architecture $\FMLP$, 
    will be a fully connected feedforward DNN (also known as a multilayer perceptron)
    equipped with the ReLU activation function 
        $\relu(x) = \max\{0,x\}$. 
    This type of network is standard in practice, however the results given here also apply with $\relu$ replaced by any continuous piecewise-linear activation function, using \citet[Corollary 3]{brown_statistical_2024}.

    For each $n\in\mathbb{N}$, 
    the graph structure of $\FMLP$ is determined by the number of hidden layers $\L$ and the number of units in each layer $\Hb=[H_{n,1},\ldots,H_{n,\L}] \in \N^{\L}$. See Figure \ref{fig:DNN_MLP} for an example of $\FMLP$.
    Each layer $l\in\{1,\ldots,\L\}$ consists of computation units or nodes, which are denoted as $\n_{l,h}$, for $h\in \{1,\ldots,H_{n,l}\}$.
    Each node, $\n_{l,h}$, is a function taking values in $\R$ which depends 
    on a 
    vector of parameters 
        $\gv_{l,h} = [\gamma_{l,h,0},\gamma_{l,h,1},\ldots,\gamma_{l,h,H_{n,l-1}}]'
        ,
        $
    and performs the computation
        \begin{equation}\label{eq:nodes}
            \n_{l,h}\left(\x  \right) 
            \coloneqq \left\{
        \begin{aligned}
        &
            \relu\bigg( 
            \sum_{i=1}^{d} \g_{1,h,i} \cdot x_i + \g_{1,h,0}
            \bigg),
            & l=1,
        \\&
            \relu\bigg( 
            \sum_{i=1}^{H_{n,l-1}} \g_{l,h,i} \cdot \n_{l-1,i}(\x) + \g_{l,h,0}
            \bigg),
            & 2\leq l \leq \L,
        \\& 
            \sum_{i=1}^{H_{n,\L}} \g_{\L+1,1,i} \cdot \n_{\L,i}(\x) + \g_{\L+1,1,0},
            &  l = \L + 1.
        \end{aligned}
        \right.
        \end{equation}
    The network consists of $\W$ parameters collected together as 
        $                
            \gV
        \coloneqq 
            \{\g_{l,h,k}\}_{\forall\, l,h,k}
        \in 
            \R^{\W}
        .
        $
    The architecture $\FMLP$ is the space of functions defined by
        \begin{equation}\label{eq:sievespace}
        \resizebox{0.92\hsize}{!}{$
            \FMLP 
        =
            \FMLPnon(\L,\Hb,\DNNBound)
        \coloneqq
            \Big\{ 
            f=
            \n_{\L+1,1}\big(\cdot \,|\,   \gV\big):
            \;  
            \gV \in \R^{\W},
            \; 
            \sup_{\x\in\Xspace}|\n_{\L+1,1}\big(\x \;|\;   \gV\big)|
            \leq \DNNBound 
            \Big\},
        $}
        \end{equation}
    where $\DNNBound\geq 2$ is the user-chosen upper bound.  
    Given an architecture $\FMLP$, 
    a particular network $f\in\FMLP$ 
    is determined by a choice of parameters $\gV$,
    and can be written as 
        $f(\cdot) = \n_{\L+1,1}\left(\cdot \;|\;   \gV\right):\Xspace\to[-\DNNBound,\DNNBound]$.

    \subsubsection{DNN Estimation}

    Following \citet{brown_statistical_2024}, we estimate $\fOT$ and $\fOY$ using the method of sieves (\citealp{grenander_abstract_1981}; \citealp{chen_chapter_2007}), 
    where the sieve spaces are constructed using DNNs.
    For the DNN architecture defined as in \eqref{eq:sievespace}, define the following DNN sieve spaces 
        \begin{equation}\label{eq:DNNs}
                \FMLPT
            \coloneqq
                \FMLPnon(\LT,\HbT,2)
        \quad
        \text{ and }
        \quad
                \FMLPY
            \coloneqq 
                \FMLPnon(\LY,\HbY,\DNNBound)   
        \end{equation}
    for  non-decreasing sequences 
        \begin{equation}\label{eq:arch_asymp}
        \begin{aligned}
            \LT \asymp \LY \asymp  \log(n)
        ,
        \quad 
            \Hi{l}^{(\ftnT)}
        \asymp 
                n^{
                    \frac{1}{2}
                    \left(\frac{d}{\smoothT+d}\right)
                }
                \log^2(n)
        ,
        \quad 
            \Hi{l}^{(\ftnY)}
        \asymp 
                n^{
                    \left(\frac{d}{\smoothY+d}\right)
                    (1/2-\DNNBoundrate)
                }
                \log^2(n)
        ,
        \end{aligned}
        \end{equation}
    for all $l\in\N$. 
    Let 
         $\fT\in\FMLPT$,
    and
         $\fY\in\FMLPY$
    be \textit{approximate sieve estimators} satisfying
        \begin{equation}\label{eq:fhat}
        \begin{aligned}
            \sampavg
            \Big(
                \Tt-\fT(\Xt)
            \Big)^2
        &\leq 
            \inf_{f\in\FMLPT} 
            \sampavg
                \Big(
                    \Tt-\ftnT(\Xt)
                \Big)^2
            + \theta_n
        ,
        \quad \text{ and }
        \\
            \sampavg
            \Big(
                \Yt-\fY(\Xt)
            \Big)^2
        &\leq 
            \inf_{f\in\FMLPY} 
            \sampavg
                \Big(
                    \Yt-\ftnT(\Xt)
                \Big)^2
            + \theta_n
        ,
        \end{aligned}
        \end{equation}
    where the ``plug-in'' error 
        $\theta_n:\Omega\to[0,\infty)$ 
    is a random variable, such that 
        $\theta_n = \op(1)$,
    and whenever feasible
    $\theta_n \coloneqq 0$.
    Note that by \citet{brown_statistical_2024} Remark 1, and Propositions 1 and 2,
    the infimums in \eqref{eq:fhat} will be measurable,
    and there will exist measurable mappings:
        $\omega\mapsto\fT$ from $\Omega$ to $\FMLPT$,
    and
        $\omega\mapsto\fY$ from $\Omega$ to $\FMLPY$. 
    Henceforth,
        $\fT$, $\fY$, will always refer to such a measurable mapping.
    To estimate $\fOT$ and $\fOY$ with $\fT$ and $\fY$ we impose the following regularity assumptions.
    \begin{assumption}\label{as:data}
    \begin{enumerate}
    \item 
        $\{\Zt\coloneqq(\Yt,\Tt,\Xt)\}_{t\in\N}$
        is a strictly stationary $\beta$-mixing process on the complete probability space $\probspace$ with
        $\beta(j)\leq \Cbeta'e^{-\Cbeta\,j } $
        for some $\Cbeta,\Cbeta'>0$. 
    \item 
        $\Yt\in\Lp{2}\probspace$, and there exists a non-decreasing sequence 
            $\{\DNNBound\}_{n\in\N}$ 
        with
            $
                B_1
            \geq
                2
            ,
            $ 
        and 
            $\DNNBound\asymp n^{\DNNBoundrate}$,
        for some $\DNNBoundrate\in[0,1/2)$
        such that
            $$
                \lim_{n\to\infty} 
                \Pr\left(\, 
                \max_{t\in\{1,\ldots,n\}}
                 |\Yt|\geq \DNNBound
                \right)
                =0
            .
            $$
        \item 
            $\Tt\in\Paramspace$, and
            $\Xt\in\Xspace$. 
        \item 
            There exist
            smoothness parameters $\smoothT\in\N$ and $\smoothY\in\N$
            for $\fOT:\Xspace\to\Paramspace$ and $\fOY:\Xspace\to\R$ 
            such that,
            for each
                $\kb,\boldsymbol{r} 
                \in 
                \big\{\N\cup\{0\}\big\}^d
                $
            where
                $\textstyle
                \sum_{j=1}^d \k_j \leq \smoothT-1$
            and
                $
                \textstyle
                \sum_{j=1}^d r_j \leq \smoothY-1$,
            the functions
                $D^{\kb}\fOT,\,$ 
                $D^{\boldsymbol{r}}\fOY$
            are continuous, 
            and 
                $
                \norm*{D^{\kb}\fOT}_\infty
                \leq 1
                ,\,
                $
                $
                \norm*{D^{\boldsymbol{r}}\fOY}_\infty
                \leq 1.
                $
    \end{enumerate}
    \end{assumption}

    Assumption \ref{as:data}(i) requires that $\{\Zt\}_{t\in\N}$ be geometrically $\beta$-mixing. 
    Notably the particular value of $\Cbeta,\Cbeta'$ will play no role in the results of this paper. 
    Stochastic-sequences that are $\beta$-mixing have been heavily studied 
    (e.g., 
    \citealp{doukhan_mixing_1994}; 
    \citealp{bradley_basic_2005}).
    For example,
    conditions to ensure stationarity and geometric $\beta$-mixing 
    for ARMA processes
    are given by \citet{mokkadem_mixing_1988},
    and for GARCH processes by
    \citet{chen_geometric_2000}, and
    \citet{carrasco_mixing_2002}. 
    Also see \citet{brown_statistical_2024} for more discussion.

    Assumptions \ref{as:data}(ii) and (iii) are standard regularity assumptions on the data. 
    Assumption \ref{as:data}(ii) will hold without $\Yt$ taking values on a compact set
    (\citealp[Assumption 1]{farrell_deep_2021}), 
    and without   
    $\E[\Yt^2|\{\Xt,\Y_{t-1}\}_{t=1}^n]$ being uniformly bounded. 
    See \citet[Proposition 3]{brown_statistical_2024} and \citet{leadbetter_extremes_1983} for examples of $\{\Yt\}_{t\in\N}$ that satisfy Assumption \ref{as:data}(ii).
    Notably, Assumption \ref{as:data}(ii) will also imply 
        $\lim_{n\to\infty}
        \Big\{
        \max_{t\in\{1,\ldots,n\}}
        \E\big[
            \Yt^2\,\mathbbm{1}_{|\Yt|\geq\DNNBound}
        \big]
        \Big\}
        =
        0
        ,
        $
    by 
    \citet[Lemma 1]{brown_statistical_2024}.

    Assumption \ref{as:data}(iv) is a H\"older smoothness condition that is widely used in the nonparametric estimation literature (e.g. \citealp{stone_optimal_1982}; \citealp{chen_sieve_1998}; \citealp{chen_chapter_2007}; \citealp{chen_optimal_2015}; \citealp{farrell_deep_2021}).
    Note that this is more general than the structural assumptions imposed in much of the DNN literature (e.g.,
        \citealp{kohler_nonparametric_2017};
        \citealp{schmidt_hieber_nonparametric_2020};
        \citealp{kohler_rate_2021}%
        ).

    Under Assumption \ref{as:data} 
    the following corollary 
    provides a rate of convergence in probability for $\fT$ and $\fY$. 
    Corollary \ref{cor:DNNinf} will follow immediately from \citet[Theorem 4]{brown_statistical_2024}.
    \begin{corollary}[\citealp{brown_statistical_2024}, Theorem 4]\label{cor:DNNinf}
        Suppose Assumption \ref{as:data} holds.
        Let 
            $\FMLPT$, $\FMLPY$ 
        be defined as in 
            \eqref{eq:DNNs}
        and
            \eqref{eq:arch_asymp}.
        Let
        $\fT\in\FMLPT$, and $\fY\in\FMLPY$ be DNN sieve estimators satisfying \eqref{eq:fhat}.
        Then,
        we have
            \begin{equation*}
            \begin{aligned}
                &
                \norm*{\fT-\fOT}_{\Lp{2}(\Px)}
                =
                    \Op\left(
                        n^{-\frac{1}{2}\left(\frac{\smoothT}{\smoothT+d}\right)}
                        \log^{6}(n)
                        +
                        \sqrt{\theta_n}
                    \right)
            ,
            \quad 
            \text{ and }
            \\&
                    \norm*{\fY-\fOY}_{\Lp{2}(\Px)}
                =
                    \Op\left(
                        n^{-\left(\frac{\smoothY}{\smoothY+d}\right)(1/2-\DNNBoundrate)}
                        \log^{6}(n)
                        +
                        \sqrt{
                        \E\big[
                            \Yt^2\,\mathbbm{1}_{\{|\Yt|\geq \DNNBound\}}
                        \big]
                        }
                        +
                        \sqrt{\theta_n}
                    \right)
            .
            \end{aligned}
            \end{equation*}
    \end{corollary}

\subsection{Asymptotic Properties of $\paramhat$}\label{sec:inf}
    The following assumption will ensure that $\paramhat$ exists with probability approaching one and that 
    Corollary \ref{cor:DNNinf}
    can be used to obtain DNN estimators of $\fOT$, $\fOY$ 
    that converge faster than $n^{-1/4}$.

    \begin{assumption}\label{as:inf}
    \begin{enumerate}
    \item 
            $
            \E\Big[\big(\Tt-\fOT(\Xt)\big)^2\Big] >0
            $,
            and $\Xt$ is continuously distributed on $\Xspace$ for each $t\in\N$.
        \item 
            The
            constant
                $\DNNBoundrate$
            and the
            smoothness parameters $\smoothT\in\N$, $\smoothY\in\N$ from Assumption \ref{as:data}
            are such that
            $$
            \frac{1}{4}
            <
            \min\left\{
                \frac{1}{2}\left(\frac{\smoothT}{\smoothT+d}\right)
            \,,\;
                \left(\frac{\smoothY}{\smoothY+d}\right)(1/2-\DNNBoundrate)
            \right\}
            ,
            $$
        and
            $
            0
            \leq
            \DNNBoundrate
            <
            \min\left\{
                    \frac{1}{4}
                    \left(
                    \frac{\smoothT-d}{\smoothT+d}
                    \right)
                \, , \;
                    \frac{1}{4}
                    \left(
                    \frac{\smoothY}{\smoothY+d/2}
                    \right)
            \right\}
            .
            $
        \vspace{10pt}
        \item 
        $\{\Yt\}_{t\in\N}$ is such that
            $\E[\,|\Yt|^{2+\delta}\,]<\infty$
        for some $\delta>0$,
        and
            $
            %
                \max_{t\in\{1,\ldots,n\}}
                \E\big[
                    \Yt^2\,\mathbbm{1}_{|\Yt|\geq\DNNBound}
                \big]
                =
                \littleo(n^{-1})
            .
            $
    \end{enumerate}
    \end{assumption}

    First, note that Assumption \ref{as:inf}(i) and Corollary \ref{cor:DNNinf} imply $\paramhat$ exists with probability approaching one.
    Next, let 
        \begin{equation}\label{eq:err}
            \err 
        \coloneqq 
            \log^{6}(n) \cdot
            \max\left\{
                n^{-\frac{1}{2}\left(\frac{\smoothT}{\smoothT+d}\right)}
            \, , \;
                n^{-\left(\frac{\smoothY}{\smoothY+d}\right)(1/2-\DNNBoundrate)}
            \right\}
        .
        \end{equation}
    Then, under Assumptions \ref{as:data} and \ref{as:inf}, Corollary \ref{cor:DNNinf} will imply 
        $\err=\littleo(n^{-1/4}),$
    and that
        $\norm*{\fT-\fOT}_{\Lp{2}(\Px)}$
    and
        $\norm*{\fY-\fOY}_{\Lp{2}(\Px)}$
    are $\Op(\err)$, whenever $\theta_n=\op(n^{-1/2})$.
%
    The upper bound on $\DNNBoundrate$ from Assumption \ref{as:inf}(ii) is equivalent to requiring $\DNNBound\err=\littleo(n^{-1/4})$.
    The feasibility of the bounds on $\DNNBoundrate$ is not an additional assumption since the first part of Assumption \ref{as:inf}(ii) implies $\smoothT>d$.  
    With this, the following theorem will use the properties of $\fT$, and $\fY$ from Corollary \ref{cor:DNNinf} to obtain $\sqrt{n}$-convergence and asymptotic normality of $\paramhat$.

    \begin{theorem}\label{thrm:inf}
    Suppose Assumptions \ref{as:data} and \ref{as:inf} hold. 
    Let $\err$ be defined as in \eqref{eq:err}.
    Let $\fT$, $\fY$ be constructed as in Corollary \ref{cor:DNNinf}, and suppose $\theta_n=\op(n^{-1/2})$.
    %
    Then, we have the following:
        \begin{enumerate}
            \item 
                $
                    \norm*{\paramhat-\paramO}
                =
                    \Op\big(n^{-1/2}\big)
                $;
            \item 
                there exists a constant $\sigma\geq0$ such that
                    $$
                        \lim_{n\to\infty}
                        \mathrm{Var}
                        \left[
                        \frac{1}{\sqrt{n}}
                        \sumin
                        \momt{\paramO}{\fOT}{\fOY}
                        \right]
                    =
                        \sigma^2
                    < 
                        \infty,
                    $$
                and if $\sigma>0$, then
                    $$
                        \sqrt{n}
                        \big(\paramhat-\paramO\big) 
                    \overset{d}{\to}
                        N
                        \left( 
                        0,
                        \;
                        \E\Big[\big(\Tt-\ftnT(\Xt)\big)^2\Big]^{-2}
                        \sigma^2
                        \right)
                    .%
\footnote{
As usual,
for a sequence of real valued random variables 
    $\{W_n\}_{n\in\N}$, 
we write 
    $
    W_n
    \overset{d}{\to}
    N
    ( 
    \mu,
    \,
    \sigma^2
    )
    $
if $W_n$ converges in distribution to a normally distributed random variable that has expected value $\mu$ and variance $\sigma^2$ as $n\to\infty$.
}
                    $$
        \end{enumerate}
    \end{theorem}

    The existence of $\sigma\geq0$ is not a new result and follows from \citet[Theorem 1.5]{bosq_1998}.
    In most settings with 
    $\E[\upsilon_t^2|\Xt,\Tt]>0$
    the requirement that
    $\sigma>0$ is fairly mild,
    and will hold whenever the covariances of 
        $
        \big\{
            {\upsilon_t}/{\big(\Tt-\ftnT(\Xt)\big)}
        \big\}_{t\in\N}
        $
    don't sum to zero.
    To see this, note that  
    by $\E[\upsilon_t|\Xt,\Tt]=0$, the definition of $\psi_t$, and Bienaym\'e's identity
        \begin{gather*}
            \mathrm{Var}
            \left[
            \frac{1}{\sqrt{n}}
            \sumin
            \momt{\paramO}{\fOT}{\fOY}
            \right]
        =
            \E
            \Bigg[
            \left(
                \frac{1}{\sqrt{n}}
                \sumin
                    \frac{\upsilon_t}
                    {\Tt-\fOT(\Xt)}
            \right)^2
            \Bigg]
        ,
        \quad
        \text{ and}
        \\
            \mathrm{Var}
            \left[
            \frac{1}{\sqrt{n}}
            \sumin
            \momt{\paramO}{\fOT}{\fOY}
            \right]
        =
            \frac{1}{n}
            \sum_{i,j=1}^n
            \mathrm{Cov}
            \big[
                \momt{\paramO}{\fOT}{\fOY}
                \psi_j\big(\paramO,\fOT,\fOY\big)
            \big]
        .
        \end{gather*}
}

\section{Conclusion}\label{sec:Conc}
This paper shows how valid inference can be obtained in a partially linear regression model under dependent data after first stage DNN estimation of infinite dimensional parameters.
Using the DNN results from \citet{brown_statistical_2024},
I show that the estimator for the finite-dimensional parameter, constructed using DNN-estimated nuisance components, achieves $\sqrt{n}$-consistency and asymptotic normality.
I do this without sample splitting, which addresses
one of the key challenges in applying machine learning techniques to econometric models with dependent data.

These results not only demonstrate the practical implications of \cite{brown_statistical_2024}, but also provide techniques that could be extended to address more complex econometric models,
such as instrumental variable models,
or more efficient estimation procedures (see e.g.,  \citealp{newey_efficient_1990}). 
This offers many promising avenues for future research that I plan to incorporate in a later version of this work.

\vx
\vx
\addappheadtotoc 
\appendix
\appendixpage
\section{Proof of Theorem \ref{thrm:inf}}\label{sec:PROOF_inf}
{
    \renewcommand{\v}[3]{
        {\mathsf{b}}
        \big(#1;#2,#3\big)
    }
%
%
%
Before proving Theorem \ref{thrm:inf} we first present three ancillary lemmas. 
These lemmas are analogous to the conditions from \citet[Theorem 1]{chen_debiased_2022} and are labeled accordingly.
    Throughout this section, we write 
        \begin{equation*}
        \begin{aligned}
        &
            \At{\ftnT} 
        = 
            \A{\Zt}{\ftnT}
        \coloneqq
            \big(\Tt-\ftnT(\Xt)\big)^2
        ,
        \quad 
        \text{ and }
        \\&
            \vt{\ftnT}{\ftnY} 
        =
            \v{\Zt}{\ftnT}{\ftnY}
        \coloneqq
            \big(
                \Tt - \ftnT(\Xt)
            \big)
            \big(
                \Yt - \ftnY(\Xt)
            \big)
        .
        \end{aligned}
        \end{equation*}
    Then, we have
        \begin{equation*}
            \paramhat
        =
            \left[\sumin \At{\ftnT} \right]^{-1}\;
            \left[\sumin \vt{\ftnT}{\ftnY} \right]
        .
        \end{equation*}

\subsection{Empirical Moment Convergence}

    \begin{lemma}\label{lem:Stab_Est}
        $$
            \sampavg\momt{\paramhat}{\fT}{\fY} = \op(n^{-1/2})
        $$
    \end{lemma}
    \begin{proof}
    Using the definition of $\paramhat$, whenever $ \sumin\A{\Zt}{\fT}\neq 0$, 
        \begin{equation*}
        \begin{aligned}
            \sampavg
        &
            \momt{\paramhat}{\fT}{\fY}
        =
            \sampavg
            \A{\Zt}{\fT}
            \,
            \paramhat
            -
            \sampavg
            \v{\Zt}{\fT}{\fY}
        \\&
        =
            \left[\sampavg\A{\Zt}{\fT}\right]
            \left[\sampavg\A{\Zt}{\fT}\right]^{-1}\;
            \left[\sampavg \v{\Zt}{\fT}{\fY}\right]
            -
            \sampavg
            \v{\Zt}{\fT}{\fY}
        \\&
        =
            \sampavg \v{\Zt}{\fT}{\fY}
            -
            \sampavg
            \v{\Zt}{\fT}{\fY}
        = 0
        .
        \end{aligned}
        \end{equation*}
    This completes the proof since 
        $ \sumin\A{\Zt}{\fT}\neq 0$ with probability approaching one under Assumption \ref{as:inf}(i) and since $\norm*{\fT-\fOT}_{\Lp{2}(\Px)}\to 0$.
    \end{proof}
\vx

\subsection{Stochastic Equicontinuity}
\begin{lemma}\label{lem:Stoch_Equi}
            \begin{equation*}
            \begin{aligned}
                \sqrt{n}
                \bigg|
                    \E\Big[
                        \At{\fT}
                        -
                        \At{\fOT}
                    \Big]
                    -
                    \sampavg
                    \Big\{
                        \At{\fT}
                        -
                        \At{\fOT}
                    \Big\}
                \bigg|
            &=
                \op(1)
            \\
                \sqrt{n}
                \bigg|
                    \E\Big[
                        \vt{\fT}{\fY}
                        -
                        \vt{\fOT}{\fOY}
                    \Big]
                    -
                    \sampavg
                    \Big\{
                        \vt{\fT}{\fY}
                        -
                        \vt{\fOT}{\fOY}
                    \Big\}
                \bigg|
            &=
                \op(1)
            \end{aligned}
            \end{equation*}
    \end{lemma}
The proof of Lemma \ref{lem:Stoch_Equi} is in Appendix \ref{sec:Stoch_Equi_Proof} below. 
Before the proof of this result, we first introduce the idea of independent blocking, commonly used when dealing with stationary $\beta$-mixing processes (e.g. \citealp{chen_sieve_1998}).

\subsubsection{Independent blocks}\label{sec:IndBlock}
\newcommand{\Ione}[1]{T_{1,#1}}%
\newcommand{\Itwo}[1]{T_{2,#1}}%
\newcommand{\IR}{T_{R}}%
\newcommand{\IZ}{\overline{\Z}}%
\newcommand{\IZt}{\IZ_t}%
\newcommand{\IX}{\overline{\X}}%
\newcommand{\IXt}{\IX_t}%
\newcommand{\dataIZ}{\{\IZt\}_{t=1}^n}%
\newcommand{\anA}{a}%
\newcommand{\anB}{b}%
    The following is from \citet[Appendix D.4]{brown_statistical_2024}.
    Let $\anA\in\N$ be such that $1\leq\anA\leq n/2$, so 
        $\anB \coloneqq \big\lfloor n/(2\anA)\big\rfloor$
    is well defined.
    Then, we can divide 
        $\data$
    into $2\anB$ blocks of length $\anA$, and the remainder into a block of length $n-2\anB\anA$,
    using the 
    index sets
        \begin{equation*}
        \begin{aligned}
            \Ione{j} 
        &\coloneqq 
            \Big\{ t\in\N:\; 2(j-1)\anA +1 \leq t \leq (2j-1)\anA  \Big\},
        \quad j=1,\ldots,\anB;
        \\
            \Itwo{j} 
        &\coloneqq 
            \Big\{ t\in\N:\; (2j-1)\anA +1 \leq t \leq 2j\anA  \Big\},
        \quad j=1,\ldots,\anB;
        \\
            \IR 
        &\coloneqq
            \Big\{ t\in\N:\; 2\anB\anA +1 \leq t \leq n  \Big\}.
        \end{aligned}
        \end{equation*}
    Using Berbee's Lemma we can  
    redefine $\data$ on a richer probability space (see \citealp[Appendix F]{brown_statistical_2024} for details)%
\footnote{We will continue to refer to the richer probability space as $\probspace$ since the extension preserves the distribution of random variables defined on the original space. 
}
    where there exists a random sequence $\dataIZ$ 
    with the following two properties:
    let
        $T,T'
        \in\{\Ione{1},\Itwo{1},\Ione{2},\ldots,\Itwo{\anB},\IR\}
        $,
    then
    (i) the block
        $\{\IZt\}_{t\in T}$
    is independent from the blocks 
        $\{\IZt\}_{t\in T'}$, 
        $\{\Zt\}_{t\in T'}$
    for any  $T'\neq T$;
    and (ii)
        $\{\IZt\}_{t\in T}$
    has the same distribution as 
        $\{\Zt\}_{t\in T}$,
    i.e. 
        $\P_{\{\IZt\}_{t\in T}}=\P_{\{\Zt\}_{t\in T}}$.
    By stationarity, all blocks of length $\anA$ are identically distributed so the sequence of blocks 
        $
            \{\IZt\}_{t\in\Ione{1}},\{\IZt\}_{t\in\Itwo{1}},\{\IZt\}_{t\in\Ione{2}},\ldots,\{\IZt\}_{t\in\Itwo{\anB}}
        $
    is i.i.d.,%
\footnote{
All blocks except $\{\IZt\}_{t\in\IR}$ are of length $\anA$, and therefore i.i.d. However,
$\dataIZ$ is not an independent sequence since elements within a single block, $\{\IZt\}_{t\in T}$, may be correlated. 
}
    and we have
        \begin{equation}\label{eq:IZT_ZT_blocks}
        \begin{aligned}
            \P_{
            \left\{\IZt: \;
            {
            t\,\in\, 
            \cup_{j=1}^{\anB} \Ione{j}
            }
            \right\}} 
        &=
            \P_{\{\IZt\}_{t\in \Ione{1}}} \times 
            \P_{\{\IZt\}_{t\in \Ione{2}}} \times 
            \cdots \times 
            \P_{\{\IZt\}_{t\in \Ione{\anB}}}
        \\&
        =
            \P_{\{\Zt\}_{t\in \Ione{1}}} \times 
            \P_{\{\Zt\}_{t\in \Ione{2}}} \times 
            \cdots \times 
            \P_{\{\Zt\}_{t\in \Ione{\anB}}}
        .
        \end{aligned}
        \end{equation}
    Next, 
    the usual $\beta$-mixing coefficient (e.g. \citealp[Definition 3.1, p.19]{dehling_empirical_2002}) can be equivalently written as (see \citealp{eberlein_weak_1984}) 
        $$
            \beta(m)
        = 
            \sup_{
                    A\times B 
                \in 
                    \sigma\left(
                    \{\Zt\}_{t=1}^k
                    \right)
                \,\otimes\,
                    \sigma\left(
                    \{\Zt\}_{t=k+m+1}^\infty
                    \right)
            }
            |\P(A \times B)- \P(A)\P(B)|.
        $$
    Hence, for $j\in \{1,...,\anB-1\}$
        \begin{equation}\label{eq:beta_bound}
        \begin{aligned}
            \beta(\anA)
        \geq
            \sup\bigg\{
            \Big|&
                \P_{\left\{\Zt:\;t\,\in \cup_{j=1}^{\anB} \Ione{j} \right\}}(A\times B)
                -
                \P_{\left\{\Zt:\;t\,\in \cup_{j=1}^k\Ione{j}\right\}}(A)\;
                \P_{\left\{\Zt:\;t\,\in \cup_{j=k+1}^{\anB}\Ione{j}\right\}}(B)
            \Big|
            : \;
        \\&
                A\times B\in 
                \sigma\big(
                \left\{\Zt:t\in \cup_{j=1}^k\Ione{j}\right\}
                \big)
            \otimes
                \sigma\big(
                \left\{\Zt:t\in \cup_{j=1}^k\Ione{j}\right\}
                \big)
            \bigg\}
        .
        \end{aligned}
        \end{equation}
    By \eqref{eq:beta_bound} the conditions for \cite{eberlein_weak_1984} Lemma 2 are satisfied. 
    So we apply this result, and use \eqref{eq:IZT_ZT_blocks},
    to obtain,
    for any measurable set $E$,
        \begin{equation*}
        \begin{aligned}
            \Big|
                \Pr\Big(
                    \Big\{\Zt: 
                    t\in
                    \cup_{j=1}^{\anB} \Ione{j}
                    \Big\}
                \in 
                    E
                \Big)
            -
                \Pr\Big(
                    \Big\{\IZt: 
                    t\in
                    \cup_{j=1}^{\anB} \Ione{j}
                    \Big\}
                \in 
                    E
                \Big)
            \Big|
        &
        \; \leq \;
            (\anB-1)\beta(\anA)
        .
        \end{aligned}
        \end{equation*}
    Then, by the triangle inequality and
        $
            \anB 
        \coloneqq 
            \big\lfloor n/(2\anA)\big\rfloor
        <
        n/(2\anA)
        +1
        ,
        $
        \begin{equation}\label{eq:beta_bound2}
        \begin{aligned}
                \Pr\Big(
                    \Big\{\Zt: 
                    t\in
                    \cup_{j=1}^{\anB} \Ione{j}
                    \Big\}
                \in 
                    E
                \Big)
        \leq
                \Pr\Big(
                    \Big\{\IZt: 
                    t\in
                    \cup_{j=1}^{\anB} \Ione{j}
                    \Big\}
                \in 
                    E
                \Big)
        +
            \frac{n\,\beta(\anA)}{2\anA}
        .
        \end{aligned}
        \end{equation}
%
%
%
%


\subsubsection{Proof of Lemma \ref{lem:Stoch_Equi}}\label{sec:Stoch_Equi_Proof}
The proof of Lemma \ref{lem:Stoch_Equi} will make use of the Rademacher complexity, defined as in \citet{brown_statistical_2024} (also see \citealp{bartlett_local_2005}).
{
\newcommand{\Rad}[1]{\mathfrak{R}_{#1}}%
\newcommand{\Radj}[2]%
    {
        \sup_{\{#2\}} 
        \frac{1}{#1} \sum_{j=1}^{#1} 
        \xi_j \normTj{f-\fO}{j}^2
    }%
    
\renewcommand{\Z}{\boldsymbol{W}}%
\renewcommand{\data}{\{\Zt\}_{t=1}^n}%
\renewcommand{\Zspace}{\R^{d_{\Z}}}%
\newcommand{\tempspace}{\mathcal{S}}%
\begin{definition}\label{def:Rad}
\textbf{\textsc{(Rademacher Complexity)}}
    For $n\in\N$, let $\data$ be random variables on $\probspace$ taking values in $\R^{d_W}$ for $d_W\in\N$.
    The Rademacher random variables, $\{\xi_t\}_{t=1}^{n}$, are i.i.d. random variables
    that are
    independent of 
        $\data$,
    and
        $\xi_t\in\{-1,1\}$ 
    where
        $\Pr(\xi_t=1)=\Pr(\xi_t=-1) = 1/2$. 
    For a pointwise-separable class of functions $\tempspace$ with elements
    $s:\Zspace\to\R$ that
    are measurable-$\borel(\Zspace)/\borel(\R)$,
    define 
        $$
            \Rad{n}\tempspace
        \coloneqq
            \sup_{s\in\tempspace}
            \frac{1}{n}\sum_{t=1}^n
            \xi_t \,
            s(\Zt)
        .
        $$
    The Rademacher complexity is 
        $\E[\Rad{n}\tempspace],$ 
    and the empirical Rademacher complexity is 
        $
            \E_{\xi}[\Rad{n}\tempspace]
        \coloneqq
            \E\big[
                \Rad{n}\tempspace
                \;\big|
                \{\Zt\}_{t=1}^n
            \big]
        $.%
\footnote{
    Note that
        $\E[\Rad{n}\tempspace]$  
    is well defined by letting $\{\xi_t\}_{t=1}^{n}$ be defined on $\probspace$ whenever $\probspace$ is rich enough, otherwise we can define $\{\xi_t\}_{t=1}^{n}$ on an auxiliary probability space 
        $(\Omega^{(\xi)},\Zsig^{(\xi)},\P^{(\xi)})$,
    and take the expectation over the product probability space
        $
            \probspace \times (\Omega^{(\xi)},\Zsig^{(\xi)},\P^{(\xi)})
        \coloneqq
            \big(
            \Omega \times S,
            \Zsig \otimes \tempspace,
            \P \times \P^{(\xi)}\big)
        .
        $
}
\end{definition}

}

{

    \begin{proof}
\newcommand{\one}{\mathbbm{1}_{nt}}%
\newcommand{\onec}{\one^{\comp}}%
\newcommand{\dif}[1]{g_{#1}}%
\newcommand{\difc}[1]{g_{#1}^{\comp}}%
    This result will follow using the ideas from the proof of \citet[Theorem 2]{brown_statistical_2024}.
    Note that $\mathsf{b}$ depends on $\Yt$ and $\ftnY\in\FMLPY$ which are not uniformly bounded for all $n$.
    Therefore,
    we will show the result for $\mathsf{b}$, and the 
    result for $\mathsf{A}$ will follow via a similar, but somewhat simpler argument.
    
    Let 
        $
            \one 
        \coloneqq 
            \mathbbm{1}\{|\Yt|\leq\DNNBound\}
        ,
        $
        and
        $
            \onec
        \coloneqq 
            \mathbbm{1}\{|\Yt|>\DNNBound\}
        ,
        $
    then define
        \begin{gather*}
            \dif{\ftnT,\ftnY}(\Zt)
        \coloneqq
            \big[\vt{\ftnT}{\ftnY}-\vt{\fOT}{\fOY}\big]
            \one
        ,
        \quad 
        \text{ and }
        \quad
            \difc{\ftnT,\ftnY}(\Zt)
        \coloneqq
            \big[\vt{\ftnT}{\ftnY}-\vt{\fOT}{\fOY}\big]
            \onec
        .
        \end{gather*} 
    With this, and the triangle inequality,
        \begin{equation}\label{eqinf:main_decomp} 
        \begin{aligned}[b]
        &
                \sqrt{n}
                \bigg|
                    \sampavg
                    \Big\{
                    \E\big[
                        \vt{\fT}{\fY}
                        -
                        \vt{\fOT}{\fOY}
                    \big]
                    -
                    \big[
                        \vt{\fT}{\fY}
                        -
                        \vt{\fOT}{\fOY}
                    \big]
                    \Big\}
                \bigg|
        \\&\leq
            \bigg|
                \frac{1}{\sqrt{n}}
                \sumin
                \Big\{
                \E\big[
                    \dif{\fT,\fY}(\Zt)
                \big]
                -
                \big[
                    \dif{\fT,\fY}(\Zt)
                \big]
                \Big\}
            \bigg|
            +
            \bigg|
                \frac{1}{\sqrt{n}}
                \sumin
                \Big\{
                \E\big[
                    \difc{\fT,\fY}(\Zt)
                \big]
                -
                \big[
                    \difc{\fT,\fY}(\Zt)
                \big]
                \Big\}
            \bigg|
        \end{aligned}
        \end{equation}

    Consider the second term of \eqref{eqinf:main_decomp}. By stationarity, for any $\delta>0$
        \begin{equation*}
        \begin{aligned}[b]
            \Pr\Bigg(
                \bigg|
        &
                    \frac{1}{\sqrt{n}}
                    \sumin
                    \Big\{
                    \E\big[
                        \difc{\fT,\fY}(\Zt)
                    \big]
                    -
                    \big[
                        \difc{\fT,\fY}(\Zt)
                    \big]
                    \Big\}
                \bigg|
            \geq
                \delta
            \Bigg)
        \\&
        \leq
            \Pr\Big(
                \sqrt{n}\,
                \E\big[
                    \,\big|\difc{\fT,\fY}(\Zt)\big|\,
                \big]
            \geq
                \delta
            \Big)
            +
            \Pr\Bigg(
                \frac{1}{\sqrt{n}}
                \sumin
                \big|\difc{\ftnT,\ftnY}(\Zt)\big|
                >0
            \Bigg)
        \end{aligned}
        \end{equation*}
    First, note that
        $|\Tt-\fT(\X)|\leq 3,\;$ 
        $|\Tt-\fOT(\X)|\leq 2<3,\;$
        $\snorm{\ftnY}\leq \DNNBound,\;$ 
    and
        $\snorm{\fOY}\leq 1 <\DNNBound$.
    Hence,
        \begin{equation*}
        \begin{aligned}
            \E\big[
                    \,\big|\difc{\fT,\fY}(\Zt)\big|\,
            \big]
        &
        \leq
            3
            \E\Big[\big|\Yt - \fY(\Xt)\big|\onec + \big|\Yt - \fOY(\Xt)\big|\onec\Big]
        \\&
        \leq
            6
            \E\Big[\big(|\Yt| + \DNNBound\big)\onec\Big]
        \leq
            12\E\big[|\Yt|\onec\big]
        = 
            \littleo(\sqrt{n})
        ,
        \end{aligned}
        \end{equation*}
    where the third inequality uses
        $\big(|\Yt| + \DNNBound\big)\onec < 2|\Yt|$ if $|\Yt|>\DNNBound$, 
        and 
        $\big(|\Yt| + \DNNBound\big)\onec = 0$ else;
    and the last equality follows from Assumption \ref{as:inf}(iii) with Jensen's inequality.
    Second,
    for any $\ftnT\in\FMLPT$, $\ftnY\in\FMLPY$
        \begin{equation*}
        \begin{aligned}
            \Pr\Bigg(
                \frac{1}{\sqrt{n}}
                \sumin
                \big|\difc{\ftnT,\ftnY}(\Zt)\big|
                >0
            \Bigg)
        &=
            \Pr\Bigg(
                \frac{1}{\sqrt{n}}
                \sumin
                \big|\vt{\ftnT}{\ftnY}-\vt{\fOT}{\fOY}\big|\onec
                >0
            \Bigg)
         \\&
         \leq
            \Pr\Bigg( 
                \frac{1}{\sqrt{n}}
                \sumin
                \mathbbm{1}_{nt}^{\comp}
                >0
            \Bigg)
        =
            \Pr\Big(
                \max_{t\in\{1,\ldots,n\}}|\Yt|>\DNNBound
            \Big)
        \to 0,\, \text{ as } n\to\infty
        ,
        \end{aligned}
        \end{equation*}
    by Assumption \ref{as:inf}(iv).
    Combining the previous three displays
        \begin{equation}\label{eqinf:trunc}
        \begin{aligned}[b]
            \Pr\Bigg(
                \bigg|
        &
                    \frac{1}{\sqrt{n}}
                    \sumin
                    \Big\{
                    \E\big[
                        \difc{\fT,\fY}(\Zt)
                    \big]
                    -
                    \big[
                        \difc{\fT,\fY}(\Zt)
                    \big]
                    \Big\}
                \bigg|
            \geq
                \delta
            \Bigg)
        =
        \littleo(1)
        .
        \end{aligned}
        \end{equation}

\newcommand{\BlockIZ}[3]{\overline{G}^{(#1)}_{#2,#3}}%
\newcommand{\BlockZ}[3]{G^{(#1)}_{#2,#3}}%

    Thus, all that remains is to address the first term of \eqref{eqinf:main_decomp}.
    We do this using the blocking from Appendix \ref{sec:IndBlock}. 
    Choose 
        $\anA\coloneqq \lceil 2\log(n) \rceil$,
    so
        $1\leq\anA\leq n/2$, 
    and 
        $\anB \coloneqq \big\lfloor n/(2\anA)\big\rfloor$ is well defined.
    Then, construct the random sequence $\dataIZ$ with the procedure described in Appendix \ref{sec:IndBlock} by dividing $\data$ into $2\anB$ blocks of length $\anA$, and the remainder into a block of length $n-2\anB\anA$, using the index sets
    $\IR$, $\Ione{j}$, $\Itwo{j}$, for $j=1,\ldots,\anB$, defined therein. 
    For 
        $m\in\{1,2\}$, 
    and
        $j\in\{1,...,\anB\}$,
    define
        $$
            \BlockZ{m}{j}{\ftnT,\ftnY}
        \coloneqq 
            \frac{1}{\anA}
            \sum_{t\in T_{m,j}}
            \dif{\ftnT,\ftnY}(\Zt)
        ,
        \quad \text{ and } \quad
            \BlockIZ{m}{j}{\ftnT,\ftnY}
        \coloneqq 
            \frac{1}{\anA}
            \sum_{t\in T_{m,j}}
            \dif{\ftnT,\ftnY}(\IZt)
        .
        $$
    With this, the first term of \eqref{eqinf:main_decomp} can be written as
        \begin{equation*}
        \resizebox{\hsize}{!}{$
        \begin{aligned}
        &
            \frac{1}{\sqrt{n}}
            \sumin
            \left\{
                \E
                \big[ 
                    \dif{\fT,\fY}(\Zt)
                \big] 
                - 
                \dif{\fT,\fY}(\Zt)
            \right\}
        \\&
        \quad
        =
            \left(\frac{\anA\anB}{\sqrt{n}}\right)
            \frac{1}{\anB}
            \sum_{j=1}^{\anB}
            \left\{
                \E\big[\BlockZ{1}{j}{\fT,\fY}\big]
                -
                \BlockZ{1}{j}{\fT,\fY}
                +
                \E\big[\BlockZ{2}{j}{\fT,\fY}\big]
                -
                \BlockZ{2}{j}{\fT,\fY}
            \right\}
        +
            \frac{1}{\sqrt{n}}
            \sum_{t\in\IR}
            \left\{
                \E
                \big[ 
                    \dif{\fT,\fY}(\Zt)
                \big] 
                - 
                \dif{\fT,\fY}(\Zt)
            \right\}
        .
        \end{aligned}   
        $}
        \end{equation*}
    Then, by stationarity, for any $\delta>0$
        \begin{equation}\label{eqinf:emp_proc_prob}
        \resizebox{0.913\hsize}{!}{$
        \begin{aligned}[b]
            &\Pr\left(
                \abs{
                \frac{1}{\sqrt{n}}
                \sumin
                \left\{
                    \E
                    \big[ 
                        \dif{\fT,\fY}(\Zt)
                    \big] 
                    - 
                        \dif{\fT,\fY}(\Zt)
                \right\}
                }
            \geq
                3\delta
            \right)
        \\&
        \leq
            2
            \Pr\left(
                \abs{
                \left(\frac{\anA\anB}{\sqrt{n}}\right)
                \frac{1}{\anB}
                \sum_{j=1}^{\anB}
                \left\{
                    \E\big[\BlockZ{1}{j}{\fT,\fY}\big]
                    -
                    \BlockZ{1}{j}{\fT,\fY}
                \right\}
                }
            \geq
                \delta
            \right)
        +
            \Pr\left(
                \abs{
                \frac{1}{\sqrt{n}}
                \sum_{t\in\IR}
                \left\{
                    \E
                    \big[ 
                        \dif{\fT,\fY}(\Zt)
                    \big] 
                    - 
                    \dif{\fT,\fY}(\Zt)
                \right\}
                }
            \geq
                \delta
            \right)
        \\&
        \coloneqq
            2\P_1 + \P_2
        .
        \end{aligned}
        $}
        \end{equation}

    Consider $\P_2$. 
    First, for any 
    $(\ftnT,\ftnY)\in \FMLPT \times \FMLPY$ 
    and all $\Zt$
        \begin{equation*}
        \begin{aligned}
                    \big|\dif{\ftnT,\ftnY}(\Zt)\big|
        &
        \leq
            3
            \Big[
                \big|\Yt - \ftnY(\Xt)\big|\one
                + 
                \big|\Yt - \fOY(\Xt)\big|\one
            \Big]
        \leq
            6
            \big(|\Yt| + \DNNBound\big)\one
        \leq
            12\DNNBound
        ,
        \end{aligned}
        \end{equation*}
    where 
    the first inequality used
        $|\Tt-\ftnT(\X)|\leq 3$ 
    and 
        $|\Tt-\fOT(\X)|\leq 2<3$;
    then
    the second inequality used
        $\snorm{\ftnY}\leq \DNNBound$, 
    and
        $\snorm{\fOY}\leq 1 <\DNNBound$.
    With this    
        $
            \big\|
            \E
            \big[ 
                \dif{\fT,\fY}(\Zt)
            \big] 
            - 
            \dif{\fT,\fY}(\Zt)
            \big\|_{\infty}
        \leq
            24\DNNBound
        $
    The cardinality of $\IR$ is 
        $(\#\IR) =  n-2\anA\anB < 2\anA$,
    since
        $\anB \coloneqq \big\lfloor n/(2\anA)\big\rfloor$
    implies
        $\anB>n/(2\anA)-1$.
    Hence,
        \begin{equation*}
        \begin{aligned}
            \abs{
            \frac{1}{\sqrt{n}}
            \sum_{t\in\IR}
            \left\{
                \E
                \big[ 
                    \dif{\fT,\fY}(\Zt)
                \big] 
                - 
                \dif{\fT,\fY}(\Zt)
            \right\}
            }
        \leq
            \frac{48\anA}{\sqrt{n}}
        =
            \littleo(1)
        ,
        \end{aligned}
        \end{equation*}
    as $n$ grows 
    since 
        $\anA\coloneqq \lceil 2\log(n) \rceil.$
    Thus, for any $\delta>0$
        \begin{equation}\label{eqinf:P2}
        \begin{aligned}
            \P_{2}
        =
            \Pr\left(
                \abs{
                \frac{1}{\sqrt{n}}
                \sum_{t\in\IR}
                \left\{
                    \E
                    \big[ 
                        \dif{\fT,\fY}(\Zt)
                    \big] 
                    - 
                    \dif{\fT,\fY}(\Zt)
                \right\}
                }
            \geq
                \delta
            \right)
        =
            \littleo(1)
        .
        \end{aligned}
        \end{equation}

\newcommand{\normT}[1]{\norm*{#1}_{\overline{T}_{1}}}%
\newcommand{\tempset}{\mathcal{H}^{(\ftnT,\ftnY)}_n(\err)}%
\newcommand{\Rad}[1]{\mathfrak{R}_{#1}}%
    Now, we address $\P_1$. 
    Define the following norm,
        \begin{equation}\label{eq:IndNorms}
        \begin{aligned}
            \normT{f}
        &\coloneqq
            \bigg(
                \frac{1}{\anB\anA}
                \sum_{j=1}^{\anB}
                \sum_{t\in \Ione{j}}\big|f(\IXt)\big|^2
            \bigg)^{1/2}
        .
        \end{aligned}
        \end{equation}
    For $\err$ as in 
    \eqref{eq:err},
    define 
        \begin{equation*}
        \begin{aligned}
            \tempset
        \coloneqq
            \Big\{
                &
                (\ftnT,\ftnY)\in \FMLPT \times \FMLPY
                \,:\,
                \norm{\ftnT-\fOT}_{\Lp{2}(\Px)}\leq C\err,\;
                \norm{\ftnY-\fOY}_{\Lp{2}(\Px)}\leq C\err,
            \\&
                \normT{\ftnT-\fOT}\leq C\err,\;
                \normT{\ftnY-\fOY}\leq C\err
            \Big\}
        \end{aligned}
        \end{equation*}
    for some $C>0$ sufficiently large such that 
        $
            \Pr\big(
            (\fT,\fY)\notin \tempset
            \big)
        =
            \littleo(1)
        $.%
\footnote{
The existence of such a $C$ sufficiently large follows from Corollary \ref{cor:DNNinf}
and the discussion 
in \citet[Appendix D.6]{brown_statistical_2024}.
To see that this 
applies here note that the 
reasoning from \citet[Appendix D.5.1]{brown_statistical_2024} can be applied
since
$\err \gtrsim r_{*}$ for $r_{*}$ defined as in 
\citet[Appendix D.5.4]{brown_statistical_2024}
and some appropriate choice of $\delta$ therein such that $\delta\to\infty$ as $n\to\infty$. 
}
    Recall 
        $\anB
        \leq
            n/(2\anA)
        ,
        $
    which implies
        $
            \left(\frac{\anA\anB}{\sqrt{n}}\right)
        \leq
            \left(\frac{\sqrt{n}}{2}\right)
        .
        $
    Then, 
    apply \eqref{eq:beta_bound2} with 
        $$%
            E=
            \left\{ 
                \abs{
                \left(\frac{\sqrt{n}}{2}\right)
                \frac{1}{\anB}
                \sum_{j=1}^{\anB}
                \left\{
                    \E\big[\BlockZ{1}{j}{\fT,\fY}\big]
                    -
                    \BlockZ{1}{j}{\fT,\fY}
                \right\}
                }
            \geq
                \delta
            \right\}
        ,
        $$
    to obtain 
        \begin{equation}\label{eqinf:Poneinit}
        \begin{aligned}[b]
            \P_1
        &\leq
            \P
            \left(
                \sup_{(\ftnT,\ftnY)\in \tempset}
                \abs{
                \left(\frac{\sqrt{n}}{2}\right)
                \frac{1}{\anB}
                \sum_{j=1}^{\anB}
                \left\{
                    \E\big[\BlockZ{1}{j}{\ftnT,\ftnY}\big]
                    -
                    \BlockZ{1}{j}{\ftnT,\ftnY}
                \right\}
                }
            \geq
                \delta
            \right)
            +
            \Pr\Big(
            (\fT,\fY)\notin \tempset
            \Big)
        \\&
        \leq
            \P
            \left(
                \sup_{(\ftnT,\ftnY)\in \tempset}
                \abs{
                \left(\frac{\sqrt{n}}{2}\right)
                \frac{1}{\anB}
                \sum_{j=1}^{\anB}
                \left\{
                    \E\big[\BlockIZ{1}{j}{\ftnT,\ftnY}\big]
                    -
                    \BlockIZ{1}{j}{\ftnT,\ftnY}
                \right\}
                }
            \geq
                \delta
            \right)
            +
            \frac{n\,\beta(\anA)}{2\anA}
            +
            \littleo(1)
        \\&
        \leq
            \P
            \left(
                \sup_{(\ftnT,\ftnY)\in \tempset}
                \abs{
                \left(\frac{\sqrt{n}}{2}\right)
                \frac{1}{\anB}
                \sum_{j=1}^{\anB}
                \left\{
                    \E\big[\BlockIZ{1}{j}{\ftnT,\ftnY}\big]
                    -
                    \BlockIZ{1}{j}{\ftnT,\ftnY}
                \right\}
                }
            \geq
                \delta
            \right)
            +
            \littleo(1)
        ,
        \end{aligned}
        \end{equation}
    by Assumption \ref{as:data}(i) and $\anA\coloneqq \lceil 2\log(n) \rceil$ imply
        $
        \frac{n\,\beta(\anA)}{2\anA}=\littleo(1)
        $.%
\footnote{
To see this holds for any 
    $\anA\gtrsim\log(n)$ 
let 
    $\delta\coloneqq \anA/\log(n)$, so $\anA= \delta \log(n)$.
Then, by Assumption \ref{as:data}(i),
    $$
        \frac{n\,\beta(\anA)}{\anA}
    \leq
        \frac{n\,\Cbeta'e^{-\Cbeta\,\anA }}{\anA}
    =
        \big(\Cbeta'\,e^{\Cbeta}\big)
        \frac{n\, e^{-\delta \log(n)}}{\anA}
    =
        \big(\Cbeta'\,e^{\Cbeta}\big)
        \frac{n^{1-\delta}}{\anA}
    =
        \big(\Cbeta'\,e^{\Cbeta}\big)
        \frac{n^{1-{\anA}/{\log(n)}}}{\anA}
    .
    $$
}
    We bound the first term on the right side with
    \citet[Theorem 2.1]{bartlett_local_2005}.
    Recall
        $
        \big|\dif{\ftnT,\ftnY}(\Zt)\big|\leq 12\DNNBound
        $
    so we have
        \begin{equation*}
        \begin{aligned}
            \max_{j\in\{1,...,\anB\}}
            \left(\frac{\sqrt{n}}{2}\right)
            \left|
                \E\big[\BlockIZ{1}{j}{\ftnT,\ftnY}\big]
                -
                \BlockIZ{1}{j}{\ftnT,\ftnY}
            \right|
        \leq
            \sqrt{n}
            \big\|\dif{\ftnT,\ftnY}(\Zt)\big\|_{\infty}
        \leq
            12\sqrt{n}\DNNBound
        .
        \end{aligned}
        \end{equation*}
    Recall from Appendix \ref{sec:IndBlock} that 
        $\P_{\{\Zt\}_{t\in\Ione{j}}}=\P_{\{\IZt\}_{t\in\Ione{j}}}, 
        \; \forall j 
        $,
    so 
    for any 
        $(\ftnT,\ftnY)\in \tempset$,
        \begin{equation*}
        \begin{aligned}
            \textrm{Var}
            \bigg[
        &
                \left(\frac{\sqrt{n}}{2}\right)
                \BlockIZ{1}{j}{\ftnT,\ftnY}
            \bigg]
        \leq
            \left(\frac{\sqrt{n}}{2}\right)^2
            \E
            \left[
                \bigg(
                \frac{1}{\anA}
                \sum_{t\in \Ione{j}}
                \dif{\ftnT,\ftnY}(\Zt)
                \bigg)^2
            \right]
        \leq
            \left(\frac{\sqrt{n}}{2}\right)^2
            \E
            \left[
                \frac{1}{\anA}
                \sum_{t\in \Ione{j}}
                \dif{\ftnT,\ftnY}(\Zt)^2
            \right]
        \\&
        =
            \left(\frac{9n}{4}\right)
            \E
            \left[
                \Big[
                    \big(\Tt - \ftnT(\Xt)\big)
                    \big(\Yt - \ftnY(\Xt)\big)
                    - 
                    \big(\Tt - \fOT(\Xt)\big)
                    \big(\Yt - \fOY(\Xt)\big)
                \Big]^2
                \one
            \right]
        \\&
        =
            \left(\frac{9n}{4}\right)
            \E
            \Bigg[
                \Big[
                    \big(\Tt - \ftnT(\Xt)\big)
                    \big(\Yt - \ftnY(\Xt)\big)
                    - 
                    \big(\Tt - \ftnT(\Xt)\big)
                    \big(\Yt - \fOY(\Xt)\big)
            \\&
            \qquad
            \qquad
            \qquad
                    +
                    \big(\Tt - \ftnT(\Xt)\big)
                    \big(\Yt - \fOY(\Xt)\big)
                    -
                    \big(\Tt - \fOT(\Xt)\big)
                    \big(\Yt - \fOY(\Xt)\big)
                \Big]^2
                \one
            \Bigg]
        \\&
        =
            \left(\frac{9n}{4}\right)
            \E
            \Bigg[
                \Big[
                    \big(\Tt - \ftnT(\Xt)\big)
                    \big(\fOY(\Xt) - \ftnY(\Xt)\big)
                    +
                    \big(\fOT(\Xt) - \ftnT(\Xt)\big)
                    \big(\Yt - \fOY(\Xt)\big)
                \Big]^2
                \one
            \Bigg]
        \\&
        \leq
            \left(\frac{9n}{4}\right)
            \E
            \Bigg[
                \Big[
                    3
                    \big|\fOY(\Xt) - \ftnY(\Xt)\big|
                    +
                    2\DNNBound
                    \big|\fOT(\Xt) - \ftnT(\Xt)\big|
                \Big]^2
            \Bigg]
        ,
        \end{aligned}
        \end{equation*}
    where the second inequality used the Cauchy-Schwarz inequality,
    the first equality used stationarity,
    and the last inequality used  
        $
            |\Tt-\ftnT(\Xt)|
        \leq
            3
        ,
        $
    and
        $
            |\Yt-\fOY(\Xt)|\one
        \leq
            \DNNBound+1
        <
            2\DNNBound
        .
        $
    Note that $3<2\DNNBound$, so we have
        \begin{equation*}
        \begin{aligned}
        \textrm{Var}
            \bigg[
        &
                \left(\frac{\sqrt{n}}{2}\right)
                \BlockIZ{1}{j}{\ftnT,\ftnY}
            \bigg]
        \leq
            \left(9\DNNBound^2 n \right)
            \E
            \Bigg[
                \Big[
                    \big|\fOY(\Xt) - \ftnY(\Xt)\big|
                    +
                    \big|\fOT(\Xt) - \ftnT(\Xt)\big|
                \Big]^2
            \Bigg]
        \\&
        =
            \left(9\DNNBound^2 n \right)
            \E
            \Big[
                    \big|\fOY(\Xt) - \ftnY(\Xt)\big|^2
                    +
                    \big|\fOT(\Xt) - \ftnT(\Xt)\big|^2
                    +
                    2
                    \big|\fOY(\Xt) - \ftnY(\Xt)\big|
                    \big|\fOT(\Xt) - \ftnT(\Xt)\big|
            \Big]
        \\&
        =
            \left(9\DNNBound^2 n \right)
            \Big[
                \|
                    \fOY - \ftnY
                \|_{\Lp{2}(\Px)}^2
                +
                \|
                    \fOT - \ftnT
                \|_{\Lp{2}(\Px)}^2
                +
                2
                \big\|
                    (\fOT - \ftnT)
                    (\fOY - \ftnY)
                \big\|_{\Lp{1}(\Px)}
            \Big]
        \\&
        \leq
            \left(9\DNNBound^2 n \right)
            \Big[
                \|
                    \fOY - \ftnY
                \|_{\Lp{2}(\Px)}^2
                +
                \|
                    \fOT - \ftnT
                \|_{\Lp{2}(\Px)}^2
                +
                2
                \|
                    \fOT - \ftnT
                \|_{\Lp{2}(\Px)}
                \|
                    \fOY - \ftnY
                \|_{\Lp{2}(\Px)}
            \Big]
        \\&
        \leq
            36\DNNBound^2\, n\, \err^2
        ,
        \end{aligned}
        \end{equation*}
    by H\"older's inequality.
    With this, since 
        $
            \{\IZt\}_{t\in\Ione{1}},
        $ 
        $
            \{\IZt\}_{t\in\Ione{2}},
        $ 
        $
            \ldots,
            \{\IZt\}_{t\in\Ione{\anB}}
        $
    is an i.i.d. sequence, we can apply \citet[Theorem 2.1]{bartlett_local_2005} 
    (with $\alpha=1/2$, and $x=\log(n)$ therein) 
    to obtain, 
\newcommand{\tempA}{\log(n)}%
        \begin{equation}\label{eqinf:Pone1}
        \resizebox{\hsize}{!}{$
        \begin{aligned}[b]
        e^{-\tempA}
        &\geq
            \P\Bigg(
                \left(\frac{\sqrt{n}}{2}\right)
                \frac{1}{\anB}
                    \sum_{j=1}^{\anB}
                    \Big\{
                        \E\big[\BlockIZ{1}{j}{\fT,\fY}\big]
                        -
                        \BlockIZ{1}{j}{\fT,\fY}
                    \Big\}
            \geq
                \left(\frac{\sqrt{n}}{2}\right)
                6
                \E_{\P^{(\xi)}}
                \Big[
                    \Rad{\anB}
                    \Big\{
                        \BlockIZ{1}{j}{\ftnT,\ftnY}:
                        (\ftnT,\ftnY)\in \tempset
                    \Big\}
                \Big] 
        \\&  
        \qquad \quad 
                +
                6\DNNBound \err
                \sqrt{
                    \frac{2n\tempA}{\anB}
                }
                +
                \frac{128\sqrt{n}\DNNBound\tempA}{\anB}
            \Bigg)
        .
        \end{aligned}
        $}        
        \end{equation}
    Now we show each of the terms in the probability bound converge to zero as $n$ grows.
    First, 
        $
            \anB
        \geq
            n/(4\anA)
        =
            n/(4\log(n))
        $
    for $n$ sufficiently large, so we have
        \begin{equation}\label{eqinf:Pone2}
        \begin{aligned}
            6\DNNBound \err
                \sqrt{
                    \frac{2n\tempA}{\anB}
                }
            +
            \frac{128\sqrt{n}\DNNBound\tempA}{\anB}
        &
        \lesssim
            \DNNBound \err \log(n)
            +
            \frac{\DNNBound\log^2(n)}{\sqrt{n}}
        =
            \littleo(1)
        .
        \end{aligned}
        \end{equation}
    Now we bound the Rademacher complexity term.
    For any 
    $(\ftnT,\ftnY)\in \FMLPT \times \FMLPY$
    and
    $(\ftnT',\ftnY')\in \FMLPT \times \FMLPY$,
        \begin{equation*}
        \begin{aligned}
            \abs{
            \BlockIZ{1}{j}{\ftnT,\ftnY}
            -
            \BlockIZ{1}{j}{\ftnT',\ftnY'}
            }
        &
        \leq
            \frac{1}{\anA}\sum_{t\in \Ione{j}}
            \Big|
                \big(\Tt - \ftnT(\Xt)\big)
                \big(\Yt - \ftnY(\Xt)\big)
                -
                \big(\Tt - \ftnT'(\Xt)\big)
                \big(\Yt - \ftnY'(\Xt)\big)
            \Big|\one
        \\&
        \leq
            \frac{1}{\anA}\sum_{t\in \Ione{j}}
            \Big|
                \big(\Tt - \ftnT(\Xt)\big)
                \big(\Yt - \ftnY(\Xt)\big)
                -
                \big(\Tt - \ftnT(\Xt)\big)
                \big(\Yt - \ftnY'(\Xt)\big)
            \\&
            \qquad
            \qquad
                +
                \big(\Tt - \ftnT(\Xt)\big)
                \big(\Yt - \ftnY'(\Xt)\big)
                -
                \big(\Tt - \ftnT'(\Xt)\big)
                \big(\Yt - \ftnY'(\Xt)\big)
            \Big|\one
        \\&
        =
            \frac{1}{\anA}\sum_{t\in \Ione{j}}
            \Big|
                \big(\Tt - \ftnT(\Xt)\big)
                \big(\ftnY(\Xt)' - \ftnY(\Xt)\big)
                +
                \big(\ftnT'(\Xt) - \ftnT(\Xt)\big)
                \big(\Yt - \ftnY'(\Xt)\big)
            \Big|\one
        \\&
        \leq
            \frac{1}{\anA}\sum_{t\in \Ione{j}}
            \Big\{
            3\big|
                \ftnY(\Xt)' - \ftnY(\Xt)
            \big|    
            +
            2\DNNBound
            \big|
                \ftnT'(\Xt) - \ftnT(\Xt)
            \big|
            \Big\}
        \\&
        \leq
            2\DNNBound\,
            \frac{1}{\anA}\sum_{t\in \Ione{j}}
            \Big\{
            \big|
                \ftnY(\Xt)' - \ftnY(\Xt)
            \big|    
            +
            \big|
                \ftnT'(\Xt) - \ftnT(\Xt)
            \big|
            \Big\}
        \\&
        \leq
            2\DNNBound\,
            \sqrt{
            \frac{1}{\anA}\sum_{t\in \Ione{j}}
            \Big\{
            \big|
                \ftnY(\Xt)' - \ftnY(\Xt)
            \big|^2    
            +
            \big|
                \ftnT'(\Xt) - \ftnT(\Xt)
            \big|^2
            \Big\}
            }
        \\&
        =
            \frac{2\DNNBound\,}{\sqrt{\anA}}
            \bigg[
            \sum_{t\in \Ione{j}}
            \Big\{
            \big|
                \big(\ftnY(\Xt)'-\fOY(\Xt)\big) - \big(\ftnY(\Xt)-\fOY(\Xt)\big)
            \big|^2
        \\&
        \qquad
        \qquad
        \qquad
        \quad
            +
            \big|
                \big(\ftnT(\Xt)'-\fOT(\Xt)\big) - \big(\ftnT(\Xt)-\fOT(\Xt)\big)
            \big|^2
            \Big\}
            \bigg]^{1/2}
        ,
        \end{aligned}
        \end{equation*}
    where the fourth inequality used
        $
            |\Tt-\ftnT(\Xt)|
        \leq
            3
        ,
        $
    and
        $
            |\Yt-\fOY(\Xt)|\one
        \leq
            \DNNBound+1
        <
            2\DNNBound
        ;
        $
    then the fifth inequality used 
        $
        2\DNNBound>3;
        $
    and the last inequality follows from the Cauchy-Schwarz inequality.
    Now we can apply \citet[Theorem 3]{maurer_vector_contraction_2016} (as in the proof of \citealp[Lemma 8]{brown_statistical_2024})
        \begin{equation*}
        \resizebox{\hsize}{!}{$
        \begin{aligned}
            \E_{\P^{(\xi)}}
            \Big[
            \Rad{\anB}
        &
                \Big\{
                    \BlockIZ{1}{j}{\ftnT,\ftnY}:
                    (\ftnT,\ftnY)\in \tempset
                \Big\}
            \Big]
        =
            \E_{\P^{(\xi)}}
            \bigg[
                \sup_{(\ftnT,\ftnY)\in \tempset}
                \frac{1}{\anB}
                \sum_{j=1}^\anB
                \xi_j
                \BlockIZ{1}{j}{\ftnT,\ftnY}
            \bigg]
        \\&
        \leq
            \frac{2\DNNBound\,}{\sqrt{\anA}\,\anB}\,
            \E_{\P^{(\xi)}}
            \bigg[
                \sup_{(\ftnT,\ftnY)\in \tempset}
                \bigg\{
                \sum_{j=1}^\anB
                \sum_{t\in\Ione{j}}
                    \xi_t
                    \big(\ftnT(\IXt)-\fOT(\IXt)\big)
            \\&
            \qquad
            \qquad
            \qquad
            \qquad
            \qquad
            \qquad
                +
                \sum_{j=1}^\anB
                \sum_{t\in\Ione{j}}
                    \xi'_t
                    \big(\ftnT(\IXt)-\fOT(\IXt)\big)
                \bigg\}
            \bigg]
        \\&
        \leq
            2\DNNBound\sqrt{\anA\,}
            \bigg(
            \E_{\P^{(\xi)}}
            \Big[
                \Rad{\anA\anB}
                \Big\{
                    \ftnT-\fOT:
                    \ftnT
                    \in \tempset
                \Big\}
            \Big]
                +
            \E_{\P^{(\xi)}}
            \Big[
                \Rad{\anA\anB}
                \Big\{
                    \ftnY-\fOY:
                    \ftnY
                    \in \tempset
                \Big\}
            \Big]
            \bigg)
        .
        \end{aligned}
        $}
        \end{equation*}
    Then, 
    using the same reasoning as display (D.22) from
        \citet[Appendix D.5.2]{brown_statistical_2024},
        \begin{equation*}
        \resizebox{\hsize}{!}{$
        \begin{aligned}
            \E_{\P^{(\xi)}}
            \Big[
            \Rad{\anB}
                \Big\{
                    \BlockIZ{1}{j}{\ftnT,\ftnY}:
                    (\ftnT,\ftnY)\in \tempset
                \Big\}
            \Big]
        &
        \leq    
            \big(24\DNNBound\sqrt{\anA\,}C\err\big)
            \sqrt{
            \frac{2\log(n)}{n}
            }
            \left(
                \sqrt{
                    \Pdim(\FMLPT)
                }
                +
                \sqrt{
                    \Pdim(\FMLPY)
                }
            \;
            \right)
        .
        \end{aligned}
        $}
        \end{equation*}
    Now, by 
    \citet[Lemma 14]{brown_statistical_2024}
    for some $C'>0$, 
        \begin{equation}\label{eqinf:Pone3}
        \resizebox{\hsize}{!}{$
        \begin{aligned}
            \E_{\P^{(\xi)}}
            \Big[
            \Rad{\anB}
                \Big\{
                    \BlockIZ{1}{j}{\ftnT,\ftnY}:
         &
                    (\ftnT,\ftnY)\in \tempset
                \Big\}
            \Big]
        \leq    
            C'\cdot
            \big(
            \DNNBound
            \sqrt{\anA\,}
            \err
            \big)
            \sqrt{
            \frac{\log^8(n)}{n}
            }
            \left(
                n^{
                    \frac{1}{2}
                    \left(\frac{d}{\smoothT+d}\right)
                }
                +
                n^{
                    \left(\frac{d}{\smoothY+d}\right)
                    (1/2-\DNNBoundrate)
                }
            \;
            \right)
        \\&
        =
            C'\cdot
            \big(
            \DNNBound
            \sqrt{\anA\,}
            \err
            \big)
            \log^4(n)
            \left(
                n^{
                    -\frac{1}{2}
                    \left(\frac{\smoothT}{\smoothT+d}\right)
                }
                +
                n^{
                    -\frac{1}{2}
                    \left(\frac{\smoothY}{\smoothY+d}\right)
                    -
                    \DNNBoundrate
                    \left(\frac{d}{\smoothY+d}\right)
                }
            \;
            \right)
        \\&
        \leq
            2C'\cdot
            \big(
            \DNNBound
            \sqrt{\anA\,}
            \err
            \big)
            \log^4(n)
            \left(
                n^{
                    -\frac{1}{4}
                }
            \right)
        \\&
        =
            \littleo\left( n^{-1/2} \right)
        ,
        \end{aligned}
        $}
        \end{equation}
    where the second inequality uses
        $
        1/4
        <
            \frac{1}{2}
            \left(\frac{\smoothT}{\smoothT+d}\right)
        $
    and
        $
        1/4
        <
            \left(\frac{\smoothY}{\smoothY+d}\right)
        $
    by Assumption \ref{as:inf}(ii),
    and the last line uses
        $
        \DNNBound\err=\littleo\left(n^{-1/4} \right)
        ,
        $
    by Assumption \ref{as:inf}(ii) and the definition of $\err$ 
    from \eqref{eq:err}.
    Thus, combining 
    \eqref{eqinf:Pone1},
    \eqref{eqinf:Pone2}, 
    and
    \eqref{eqinf:Pone3},
        \begin{equation*}
        \begin{aligned}[b]
        e^{-\tempA}
        &\geq
            \P\Bigg(
                \left(\frac{\sqrt{n}}{2}\right)
                \frac{1}{\anB}
                    \sum_{j=1}^{\anB}
                    \Big\{
                        \E\big[\BlockIZ{1}{j}{\fT,\fY}\big]
                        -
                        \BlockIZ{1}{j}{\fT,\fY}
                    \Big\}
            \geq
            \littleo(1)
            \Bigg)
        .
        \end{aligned}
        \end{equation*}
    With \eqref{eqinf:Poneinit} this implies $\P_1=\littleo(1)$.
    Combining this with 
    \eqref{eqinf:main_decomp},
    \eqref{eqinf:trunc},
    \eqref{eqinf:emp_proc_prob},
    and
    \eqref{eqinf:P2},
    implies
        \begin{equation*}
        \begin{aligned}
            \sqrt{n}
            \bigg|
                \E\Big[
                    \vt{\fT}{\fY}
                    -
                    \vt{\fOT}{\fOY}
                \Big]
                -
                \sampavg
                \Big\{
                    \vt{\fT}{\fY}
                    -
                    \vt{\fOT}{\fOY}
                \Big\}
            \bigg|
        &=
            \op(1)
        .
        \end{aligned}
        \end{equation*}
    The proof is complete since
        $
            \sqrt{n}
            \Big|
                \E\Big[
                    \At{\fT}
                    -
                    \At{\fOT}
                \Big]
                -
                \sampavg
                \Big\{
                    \At{\fT}
                    -
                    \At{\fOT}
                \Big\}
            \Big|
        =
            \op(1),
        $
    follows by a similar argument. 
\end{proof}
}
\vx

\subsection{Neyman Orthogonality}
{
\newcommand{\Ftn}{F}%
\newcommand{\F}{\hat{\Ftn}}%
\newcommand{\FO}{\Ftn_{0}}%
\newcommand{\momF}[3]{\psi\big(#1;#2,#3\big)}
\newcommand{\MomF}[2]{
\E\Big[{\psi}\big(#1,#2\big)\Big]
}
\renewcommand{\Mom}[3]{\E\Big[{\psi}\big(#1,#2,#3\big)\Big]}
\begin{lemma}\label{lem:Ney_Orth}
    For any $\ftnT,\ftnY \in\Lp{\infty}(\Px)$ 
    such that 
    $\norm*{\ftnT}_\infty\leq 2$,
    $\norm*{\ftnY}_\infty\leq \DNNBound$   
        \begin{equation*}
        \begin{aligned}
            \Mom{\paramO}{\fOT}{\fOY}
            -
            \Mom{\paramO}{\fT}{\fY}
        &=
            \op(n^{-1/2})
        .
        \end{aligned}
        \end{equation*}
\end{lemma}
\begin{proof}
    For arbitrary $\ftnT,\ftnY \in\Lp{\infty}(\Px)$ 
    such that 
        $\norm*{\ftnT}_\infty\leq 2$,
        $\norm*{\ftnY}_\infty\leq \DNNBound$
    and $\lambda\in\R$ define
        $$
            \Ftn(\lambda)
        \coloneqq
            (\fOT,\fOY)^{\tr}
            +
            \lambda
            \big[
            (\ftnT,\ftnY)^{\tr}
            -
            (\fOT,\fOY)^{\tr}
            \big]
        .
        $$
    With this, we write 
        $$
            \MomF{\paramO}{\Ftn(0)}
        =
            \Mom{\paramO}{\fOT}{\fOY}
        ,
        \quad
        \text{ and }
        \quad
            \MomF{\paramO}{\Ftn(1)}
        =
            \Mom{\paramO}{\ftnT}{\ftnY}
        .
        $$
    Applying Taylor's Theorem to 
        $\MomF{\paramO}{\Ftn(\lambda)}$
    centered at $\lambda=0$
        \begin{equation*}
        \begin{aligned}
            \Mom{\paramO}{\ftnT}{\ftnY}
        &=
            \MomF{\paramO}{\Ftn(0)}
            +
            \bigg[
                \frac{d}{d \lambda}
                \MomF{\paramO}{\Ftn(\lambda)}
            \bigg]_{\lambda=0}
            +
            \frac{1}{2}
            \int_{0}^{1}
                \bigg[
                \frac{d^2}{d \lambda^2}
                \MomF{\paramO}{\Ftn(\lambda)}
                \bigg]
            d\lambda
        \end{aligned}
        \end{equation*}
    For the first-order derivatives of $\psi$
        \begin{equation*}
        \begin{aligned}
                \frac{\partial}{\partial \lambda}
                \A{\Zt}{\fOT+\lambda(\ftnT-\fOT)}
        &=
            2\big(\ftnT(\Xt)-\fOT(\Xt)\big)
            \Big(
                \Tt-\fOT(\Xt)
                +
                \lambda
                \big(\ftnT(\Xt)-\fOT(\Xt)\big)
            \Big)
    ,
    \\
                \frac{\partial}{\partial \lambda}
                \mathsf{b}\big(\Zt;\Ftn(\lambda)\big)
        &=
                \big(\ftnT(\Xt)-\fOT(\Xt)\big)
                \Big(
                    \Yt-\fOY(\Xt)
                    +
                    \lambda
                    \big(\ftnY(\Xt)-\fOY(\Xt)\big)
                \Big)
        \\&
        \quad
            +
                \big(\ftnY(\Xt)-\fOY(\Xt)\big)
                \Big(
                    \Tt-\fOT(\Xt)
                    +
                    \lambda
                    \big(\ftnT(\Xt)-\fOT(\Xt)\big)
                \Big)
        .
        \end{aligned}
        \end{equation*}
    For the second-order derivatives of $\psi$
        \begin{equation*}
        \begin{aligned}
                \frac{\partial^2}{\partial \lambda^2}
                \A{\Zt}{\fOT+\lambda(\ftnT-\fOT)}
        &=
            2\big(\ftnT(\Xt)-\fOT(\Xt)\big)^2
        ,
    \\
                \frac{\partial^2}{\partial \lambda^2}
                \mathsf{b}\big(\Zt;\Ftn(\lambda)\big)
        &=
                2\big(\ftnT(\Xt)-\fOT(\Xt)\big)
                \big(\ftnY(\Xt)-\fOY(\Xt)\big)
        ,
        \end{aligned}
        \end{equation*}
    Hence
        \begin{equation*}
        \resizebox{\hsize}{!}{$
        \begin{aligned}
            \bigg[
                \frac{\partial}{\partial \lambda}
                \momF{\Zt}{\paramO}{\Ftn(\lambda)} 
            \bigg]_{\lambda=0}
        &=
            2
                \big(\ftnT(\Xt)-\fOT(\Xt)\big)
                \big(\Tt-\fOT(\Xt)\big)
        \\&
        \;
            -
                \big(\ftnT(\Xt)-\fOT(\Xt)\big)
                \big(\Yt-\fOY(\Xt)\big)
            -
                \big(\ftnY(\Xt)-\fOY(\Xt)\big)
                \big(\Tt-\fOT(\Xt)\big)
        .
        \end{aligned}
        $}
        \end{equation*}
    Note that for any $\delta>0$, 
    and any 
        $\lambda\in (-\delta,1+\delta)$
    the function 
        $\MomF{\paramO}{\Ftn(\lambda)}$ is continuously differentiable infinitely many times
    and its derivatives are elements of $\Lp{2}(\Pz)$ since $\ftnT,\ftnY,\fOT,\fOY$ are bounded and $\Yt\in\Lp{2}$.
    Therefore,
    a measure-theoreic version of Leibniz's Rule (e.g. \citealp[Theorem 7.5.19, p.405]{corbae_introduction_2009}) can be applied to show $\psi$ satisfies a Neyman Orthogonality condition,
        \begin{equation*}
        \resizebox{\hsize}{!}{$
        \begin{aligned}
            \bigg[
        &
                \frac{d}{d \lambda}
                \MomF{\paramO}{\Ftn(\lambda)}
            \bigg]_{\lambda=0}
        \\&
        =
            2
            \E\bigg[
                \big(\ftnT(\Xt)-\fOT(\Xt)\big)
                \big(\Tt-\fOT(\Xt)\big)
            \bigg]
            -
            \E\bigg[
                \big(\ftnT(\Xt)-\fOT(\Xt)\big)
                \big(\Yt-\fOY(\Xt)\big)
            \bigg]
        \\&
        \quad
            -
            \E\bigg[
                \big(\ftnY(\Xt)-\fOY(\Xt)\big)
                \big(\Tt-\fOT(\Xt)\big)
            \bigg]
        \\&
        =
            2
            \E\bigg[
                \big(\ftnT(\Xt)-\fOT(\Xt)\big)
                \big(\E[\Tt|\Xt]-\fOT(\Xt)\big)
            \bigg]
            -
            \E\bigg[
                \big(\ftnT(\Xt)-\fOT(\Xt)\big)
                \big(\E[\Yt|\Xt]-\fOY(\Xt)\big)
            \bigg]
        \\&
        \quad
            -
            \E\bigg[
                \big(\ftnY(\Xt)-\fOY(\Xt)\big)
                \big(\E[\Tt|\Xt]-\fOT(\Xt)\big)
            \bigg]
        \\&
        = 0
        ,
        \end{aligned}
        $}
        \end{equation*}
    by iterated expectations and the definitions of $\fOT$, $\fOY$. Applying this to the Taylor series expansion from before, and by similar arguments we can apply Leibnez's rule again and Fubini's theorem to the remainder term to obtain 
        \begin{equation*}
        \resizebox{\hsize}{!}{$
        \begin{aligned}
            \Mom{\paramO}{\ftnT}{\ftnY}
        &=
            \MomF{\paramO}{\Ftn(0)}
            +
            \frac{1}{2}
            \int_{0}^{1}
                \bigg[
                \frac{d^2}{d \lambda^2}
                \MomF{\paramO}{\Ftn(\lambda)}
                \bigg]
            d\lambda
        \\&
        =
            \MomF{\paramO}{\Ftn(0)}
            +
            \E\Big[
                    \big(\ftnT(\Xt)-\fOT(\Xt)\big)^2
                    -
                    \big(\ftnT(\Xt)-\fOT(\Xt)\big)
                    \big(\ftnY(\Xt)-\fOY(\Xt)\big)
            \Big]
        \\&
        \leq
            \MomF{\paramO}{\Ftn(0)}
            +
            \norm*{\ftnT-\fOT}_{\Lp{2}(\Px)}^2
            +
            \norm*{\ftnT-\fOT}_{\Lp{2}(\Px)}
            \norm*{\ftnY-\fOY}_{\Lp{2}(\Px)}
        \\&
        =
            \MomF{\paramO}{\Ftn(0)}
            +
            \Op(\err^2)
        ,
        \end{aligned}
        $}
        \end{equation*}
    for $\err$ defined as in \eqref{eq:err}.
    The desired result follows because 
        $\MomF{\paramO}{\Ftn(0)}
        =
            \Mom{\paramO}{\fOT}{\fOY}$
    by definition, and
        $\err = \littleo\big(n^{-1/4}\big)$
    by Assumption \ref{as:inf}.
\end{proof}

}

\vx
\vx

\newcommand{\emp}{{U}_{n}}%
\begin{proof}[Proof of Theorem \ref{thrm:inf}]
This proof is similar to \citet[Theorem 1]{chen_debiased_2022}, but specifically tailored to settings with dependent data. 
First,
        \begin{equation*}
    \begin{aligned}
        \E\Big[
        \big(\At{\fT}-\At{\fOT}\big)
        \Big]
    &=
        \E\Big[
        \big(\Tt-\fT(\Xt)\big)^2
        -
        \big(\Tt-\fOT(\Xt)\big)^2
        \Big]
    \\&
    =
        \E\Big[
        \fT(\Xt)^2
        -
        \fOT(\Xt)^2
        +
        2\Tt\big(
            \fOT(\Xt)
            -
            \fT(\Xt)
        \big)
        \Big]
    \\&
    =
        \E\Big[
        \big(
            \fOT(\Xt)
            -
            \fT(\Xt)
        \big)
        \big(
            \fOT(\Xt)
            +
            \fT(\Xt)
            +
            2\Tt
        \big)
        \Big]
    \\&
    \leq
        5\,
        \E\Big[
        \big|
            \fOT(\Xt)
            -
            \fT(\Xt)
        \big|
        \Big]
    \leq
        5\,
        \norm*{\fT-\fOT}_{\Lp{2}(\Px)}
    =
    \Op(\err)
    .
    \end{aligned}
    \end{equation*}
With this,
    \begin{equation}\label{eqinf:proofstart1}
    \begin{aligned}[b]
        \E\big[\At{\fT}\big]\big(\paramhat-\paramO\big) 
    &=
        \E\big[\At{\fOT}\big]    
            \big(\paramhat-\paramO\big) 
            +
            \E
           \big[\At{\fT} - \At{\fOT}\big]
           \big(\paramhat-\paramO\big) 
    \\&=
            \E\big[\At{\fOT}\big]    
            \big(\paramhat-\paramO\big) 
        +
        \Op(\err)
        \bigO\big( |\paramhat-\paramO| \big)
    \\&=
        \E\big[\At{\fOT}\big]    
            \big(\paramhat-\paramO\big) 
        +
        \op\big( |\paramhat-\paramO| \big)
    .
    \end{aligned}
    \end{equation}
Define the following notation
    \begin{gather*}
        \Mom{\param}{\ftnT}{\ftnY}
    \coloneqq
        \E\big[\At{\ftnT}\big]\param
        -
        \E\big[\vt{\ftnT}{\ftnY}\big]    
    ,
    \qquad
        \Momn{\param}{\ftnT}{\ftnY}
    \coloneqq
        \sampavg
        \big\{\At{\ftnT}\param
        -
        \vt{\ftnT}{\ftnY}\big\}
    ,
    \\
    \text{and} \quad 
        \emp(\param,\ftnT,\ftnY) 
    \coloneqq 
        \Mom{\param}{\ftnT}{\ftnY}
        -
        \Momn{\param}{\ftnT}{\ftnY}
        .
    \end{gather*}
Then, 
    \begin{equation}\label{eqinf:proofstart2}
    \begin{aligned}[b]
        \E\big[\At{\fT}\big]\big(\paramhat-\paramO\big) 
    &=
            \Mom{\paramhat}{\fT}{\fY}
            -
            \Mom{\paramO}{\fT}{\fY}
    \\&
    =
        \emp\big(\paramhat,\fT,\fY\big) 
        -
        \Mom{\paramO}{\fT}{\fY}
        +
        \Momn{\paramhat}{\fT}{\fY}
    \\&
    =
        \emp\big(\paramhat,\fT,\fY\big) 
        +
        \Mom{\paramO}{\fOT}{\fOY}
        -
        \Mom{\paramO}{\fT}{\fY}
        +
        \Momn{\paramhat}{\fT}{\fY}
    \\&
    =
        \emp\big(\paramhat,\fT,\fY\big) 
        +
        \op(n^{-1/2})
    ,
    \end{aligned}
    \end{equation}
where the third equality uses 
    $
    \Mom{\paramO}{\fOT}{\fOY}
    =
        \E\big[\At{\ftnT}\paramO
        -
        \vt{\ftnT}{\ftnY}\big] 
    =
    0
    $,
and the last equality uses 
    Lemma \ref{lem:Stab_Est} and Lemma \ref{lem:Ney_Orth}.
Combining \eqref{eqinf:proofstart1} and \eqref{eqinf:proofstart2}
    \begin{equation}\label{eqinf:predecomp}
    \begin{aligned}[b]
        \E\big[\At{\fOT}\big]    
            \big(\paramhat-\paramO\big) 
        +
        \op\big( |\paramhat-\paramO| \big)
    =
        \emp\big(\paramhat,\fT,\fY\big) 
        +
        \op(n^{-1/2})
    .
    \end{aligned}
    \end{equation}
Now we decompose $\emp\big(\paramhat,\fT,\fY\big) $ into an asymptotically normal component and components that converge to zero in probability at a suitable rate,
    \begin{equation}\label{eqinf:EmpProc_Decomp}
    \resizebox{.9\hsize}{!}{$
    \begin{aligned}
        &
        \emp\big(\paramhat,\fT,\fY\big) 
    \\
    &
    =
        \emp\big(\paramO,\fOT,\fOY\big) 
        +
        \Big[
            \emp\big(\paramhat,\fT,\fY\big)
            -
            \emp\big(\paramO,\fT,\fY\big)
        \Big]
        +
        \Big[
            \emp\big(\paramO,\fT,\fY\big)
            -
            \emp\big(\paramO,\fOT,\fOY\big) 
        \Big]
    .
    \end{aligned}
    $}
    \end{equation}
For the middle term of \eqref{eqinf:EmpProc_Decomp}
    \begin{equation*}
    \begin{aligned}
        \emp\big(\paramhat,\fT,\fY\big)
        -
        \emp\big(\paramO,\fT,\fY\big)
    &=
        \Mom{\paramhat}{\fT}{\fY}
        -
        \Momn{\paramhat}{\fT}{\fY}
        -
        \Mom{\paramO}{\fT}{\fY}
        +
        \Momn{\paramO}{\fT}{\fY}
    \\&
    =
        \big(\paramhat-\paramO\big) 
        \bigg(
            \E\big[\At{\fT}\big]
            -
            \sampavg
            \At{\fT}
        \bigg)
    .
    \end{aligned}
    \end{equation*}
Then,
by the triangle inequality
    \begin{equation*}
    \begin{aligned}
        \bigg|
    &
            \E\big[\At{\fT}\big]
            -
            \sampavg
            \At{\fT}
        \bigg|
    \\&
    \leq
        \bigg|
            \E\big[\At{\fOT}\big]
            -
            \sampavg
            \At{\fOT}
        \bigg|
        +
        \bigg|
            \E\Big[
                \At{\fT}
                -
                \At{\fOT}
            \Big]
            -
            \sampavg
            \Big\{
                \At{\fT}
                -
                \At{\fOT}
            \Big\}
        \bigg|
    \\&
    =
        \op\Big(n^{-1/2} \log_2(n)\log(n)\Big)
        +
        \op(n^{-1/2})
    ,
    \end{aligned}
    \end{equation*}
    where the last line follows by applying 
    Lemma \ref{lem:Stoch_Equi} to the second term, 
    and 
    \citet[Theorem 1.6, p.35]{bosq_1998} to the first term, 
    which can be applied since 
    $|\At{\fOT}|=\big|\Tt-\fOT(\Xt)\big|^2\leq 4$,
    and
    $\At{\fOT}$ 
    inherits the mixing properties of $\data$
    by \citet[Theorem 15.1]{davidson_stochastic_2022}.
    Combining the previous two displays,
        \begin{equation*}
        \begin{aligned}
            \emp\big(\paramhat,\fT,\fY\big)
            -
            \emp\big(\paramO,\fT,\fY\big)
        &=
            \op\big( |\paramhat-\paramO| \big)
        .
        \end{aligned}
        \end{equation*}
    For the last term of \eqref{eqinf:EmpProc_Decomp},
        \begin{equation*}
        \resizebox{\hsize}{!}{$
        \begin{aligned}
            \emp\big(\paramO,\fT,\fY\big)
            -
            \emp\big(\paramO,\fOT,\fOY\big)
        &\leq
            \bigg|
                \E\Big[
                    \At{\fT}
                    -
                    \At{\fOT}
                \Big]
                -
                \sampavg
                \Big\{
                    \At{\fT}
                    -
                    \At{\fOT}
                \Big\}
            \bigg|\paramO
        \\&
        \;\;\;
            +
            \bigg|
                \E\Big[
                    \vt{\fT}{\fY}
                    -
                    \vt{\fOT}{\fOY}
                \Big]
                -
                \sampavg
                \Big\{
                    \vt{\fT}{\fY}
                    -
                    \vt{\fOT}{\fOY}
                \Big\}
            \bigg|
        \\&
        =
        \op(n^{-1/2})
        ,
        \end{aligned}
        $}
        \end{equation*}
    by Lemma \ref{lem:Stoch_Equi}.
    Applying the previous two displays to \eqref{eqinf:EmpProc_Decomp} and plugging this into \eqref{eqinf:predecomp}
        \begin{equation}\label{eqinf:finalbound}
        \begin{aligned}[b]
            \E\big[\At{\fOT}\big]    
                \big(\paramhat-\paramO\big) 
            +
            \op\big( |\paramhat-\paramO| \big)
        =
            \emp\big(\paramO,\fOT,\fOY\big)  
            +
            \op(n^{-1/2})
        .
        \end{aligned}
        \end{equation}

    Note that $\emp\big(\paramO,\fOT,\fOY\big)$ is a zero-mean process,
    and
        $        
            \emp\big(\paramO,\fOT,\fOY\big)
        =
            -
            \sampavg\momt{\paramO}{\fOT}{\fOY}
        .
        $
    Then, by \citet[Theorem 1.5, p.34]{bosq_1998}, for some constant $\sigma^2\geq0$
        \begin{equation*}
        \begin{aligned}
            \textrm{Var}\Bigg[
                \frac{1}{\sqrt{n}}
                \sumin
                \momt{\paramO}{\fOT}{\fOY}
            \Bigg]
        &=
            \textrm{Var}\Big[\sqrt{n}\,\emp\big(\paramO,\fOT,\fOY\big)\Big]
        =
            \E
            \bigg[\Big(\sqrt{n}\,\emp\big(\paramO,\fOT,\fOY\big)\Big)^2\bigg]
        \\&=
            \Big\|
            \sqrt{n}\,\emp\big(\paramO,\fOT,\fOY\big)
            \Big\|_{\Lp{2}}^2
        \to 
            \sigma^2
        ,
        \end{aligned}
        \end{equation*}
    as $n\to \infty$. 
    By Markov's inequality, and H\"older's inequality, for any $c>0$,
        \begin{equation*}
        \begin{aligned}
            \Pr\Big(
                \big|
                    \sqrt{n}\,\emp\big(\paramO,\fOT,\fOY\big)
                \big|
            \geq
                c
            \Big)
        &\leq
            \frac{1}{c}\,
            \Big\|
            \sqrt{n}\,\emp\big(\paramO,\fOT,\fOY\big)
            \Big\|_{\Lp{1}}
        \leq
            \frac{1}{c}\,
            \Big\|
            \sqrt{n}\,\emp\big(\paramO,\fOT,\fOY\big)
            \Big\|_{\Lp{2}}
        \\&
        =
            \frac{1}{c}\,
            \sqrt{\textrm{Var}\Big[\sqrt{n}\,\emp\big(\paramO,\fOT,\fOY\big)\Big]}
        \to 
            \frac{\sigma}{c}
        ,
        \end{aligned}
        \end{equation*}
    as $n\to\infty$.
    Thus, 
        $
        \emp\big(\paramO,\fOT,\fOY\big) = \Op(n^{-1/2})
        .
        $
    Applying this to \eqref{eqinf:finalbound}, since $\E\big[\At{\fOT}\big] >0$ is a constant,
        \begin{equation*}
        \begin{aligned}[b]
        &
        \E\big[\At{\fOT}\big]    
                \big(\paramhat-\paramO\big) 
            +
            \op\big( |\paramhat-\paramO| \big)
        =
            \Op(n^{-1/2})
        \quad
        \implies
        \quad
            \big(\paramhat-\paramO\big)
            =
            \Op(n^{-1/2})
        ,
        \end{aligned}
        \end{equation*}
    which proves result (i).
    
    With this, we can write \eqref{eqinf:finalbound} as
        \begin{equation*}
        \begin{aligned}[b]
            \sqrt{n}
            \big(\paramhat-\paramO\big) 
        =
            \sqrt{n}
            \,
            \E\big[\At{\fOT}\big]^{-1}
            \,
            \emp\big(\paramO,\fOT,\fOY\big)  
            +
            \op(1)
        .
        \end{aligned}
        \end{equation*}
    Result (ii) follows by applying a central limit theorem to the first term.
    If $\sigma>0$, then the conditions for \citet[Theorem 1.7]{bosq_1998} are met by Assumption \ref{as:inf}(i)(ii), since 
        $\emp\big(\paramO,\fOT,\fOY\big)$
    inherits the mixing properties of $\data$
    by \citet[Theorem 15.1]{davidson_stochastic_2022}.
    Then, this result implies
        $$
        \frac{\sqrt{n}\,\emp\big(\paramO,\fOT,\fOY\big)}{\sigma}
        \overset{d}{\to}
        N(0,1)
        .
        $$
\end{proof}

}

\singlespacing
\bibliographystyle{ACM-Reference-Format}
\bibliography{References}


\begin{thebibliography}{35}


\ifx \showCODEN    \undefined \def \showCODEN     #1{\unskip}     \fi
\ifx \showDOI      \undefined \def \showDOI       #1{#1}\fi
\ifx \showISBNx    \undefined \def \showISBNx     #1{\unskip}     \fi
\ifx \showISBNxiii \undefined \def \showISBNxiii  #1{\unskip}     \fi
\ifx \showISSN     \undefined \def \showISSN      #1{\unskip}     \fi
\ifx \showLCCN     \undefined \def \showLCCN      #1{\unskip}     \fi
\ifx \shownote     \undefined \def \shownote      #1{#1}          \fi
\ifx \showarticletitle \undefined \def \showarticletitle #1{#1}   \fi
\ifx \showURL      \undefined \def \showURL       {\relax}        \fi
\providecommand\bibfield[2]{#2}
\providecommand\bibinfo[2]{#2}
\providecommand\natexlab[1]{#1}
\providecommand\showeprint[2][]{arXiv:#2}

\bibitem[Bartlett et~al\mbox{.}(2005)]%
        {bartlett_local_2005}
\bibfield{author}{\bibinfo{person}{Peter~L. Bartlett}, \bibinfo{person}{Olivier Bousquet}, {and} \bibinfo{person}{Shahar Mendelson}.} \bibinfo{year}{2005}\natexlab{}.
\newblock \showarticletitle{Local {Rademacher} complexities}.
\newblock \bibinfo{journal}{\emph{The Annals of Statistics}} \bibinfo{volume}{33}, \bibinfo{number}{4} (\bibinfo{date}{Aug.} \bibinfo{year}{2005}).
\newblock
\showISSN{0090-5364}
\urldef\tempurl%
\url{https://doi.org/10.1214/009053605000000282}
\showDOI{\tempurl}
\newblock
\shownote{arXiv:math/0508275}.


\bibitem[Bosq(1998)]%
        {bosq_1998}
\bibfield{author}{\bibinfo{person}{D. Bosq}.} \bibinfo{year}{1998}\natexlab{}.
\newblock \bibinfo{booktitle}{\emph{Nonparametric {Statistics} for {Stochastic} {Processes}}}. \bibinfo{series}{Lecture {Notes} in {Statistics}}, Vol.~\bibinfo{volume}{110}.
\newblock \bibinfo{publisher}{Springer}, \bibinfo{address}{New York, NY}.
\newblock
\showISBNx{978-0-387-98590-9 978-1-4612-1718-3}
\urldef\tempurl%
\url{https://doi.org/10.1007/978-1-4612-1718-3}
\showDOI{\tempurl}


\bibitem[Bradley(2005)]%
        {bradley_basic_2005}
\bibfield{author}{\bibinfo{person}{Richard~C. Bradley}.} \bibinfo{year}{2005}\natexlab{}.
\newblock \showarticletitle{Basic {Properties} of {Strong} {Mixing} {Conditions}. {A} {Survey} and {Some} {Open} {Questions}}.
\newblock \bibinfo{journal}{\emph{Probability Surveys}}  \bibinfo{volume}{2} (\bibinfo{date}{Jan.} \bibinfo{year}{2005}).
\newblock
\showISSN{1549-5787}
\urldef\tempurl%
\url{https://doi.org/10.1214/154957805100000104}
\showDOI{\tempurl}
\newblock
\shownote{arXiv:math/0511078}.


\bibitem[Brown(2024)]%
        {brown_statistical_2024}
\bibfield{author}{\bibinfo{person}{Chad Brown}.} \bibinfo{year}{2024}\natexlab{}.
\newblock \bibinfo{title}{Statistical {Properties} of {Deep} {Neural} {Networks} with {Dependent} {Data}}.
\newblock
\newblock
\urldef\tempurl%
\url{https://doi.org/10.48550/arXiv.2410.11113}
\showDOI{\tempurl}
\newblock
\shownote{arXiv:2410.11113}.


\bibitem[Carrasco and Chen(2002)]%
        {carrasco_mixing_2002}
\bibfield{author}{\bibinfo{person}{Marine Carrasco} {and} \bibinfo{person}{Xiaohong Chen}.} \bibinfo{year}{2002}\natexlab{}.
\newblock \showarticletitle{{MIXING} {AND} {MOMENT} {PROPERTIES} {OF} {VARIOUS} {GARCH} {AND} {STOCHASTIC} {VOLATILITY} {MODELS}}.
\newblock \bibinfo{journal}{\emph{Econometric Theory}} \bibinfo{volume}{18}, \bibinfo{number}{1} (\bibinfo{date}{Feb.} \bibinfo{year}{2002}), \bibinfo{pages}{17--39}.
\newblock
\showISSN{0266-4666, 1469-4360}
\urldef\tempurl%
\url{https://doi.org/10.1017/S0266466602181023}
\showDOI{\tempurl}


\bibitem[Chen and Chen(2000)]%
        {chen_geometric_2000}
\bibfield{author}{\bibinfo{person}{Min Chen} {and} \bibinfo{person}{Gemai Chen}.} \bibinfo{year}{2000}\natexlab{}.
\newblock \showarticletitle{Geometric {Ergodicity} of {Nonlinear} {Autoregressive} {Models} with {Changing} {Conditional} {Variances}}.
\newblock \bibinfo{journal}{\emph{The Canadian Journal of Statistics / La Revue Canadienne de Statistique}} \bibinfo{volume}{28}, \bibinfo{number}{3} (\bibinfo{year}{2000}), \bibinfo{pages}{605--613}.
\newblock
\showISSN{0319-5724}
\urldef\tempurl%
\url{https://doi.org/10.2307/3315968}
\showDOI{\tempurl}
\newblock
\shownote{Publisher: [Statistical Society of Canada, Wiley]}.


\bibitem[Chen et~al\mbox{.}(2022)]%
        {chen_debiased_2022}
\bibfield{author}{\bibinfo{person}{Qizhao Chen}, \bibinfo{person}{Vasilis Syrgkanis}, {and} \bibinfo{person}{Morgane Austern}.} \bibinfo{year}{2022}\natexlab{}.
\newblock \bibinfo{title}{Debiased {Machine} {Learning} without {Sample}-{Splitting} for {Stable} {Estimators}}.
\newblock
\newblock
\urldef\tempurl%
\url{http://arxiv.org/abs/2206.01825}
\showURL{%
\tempurl}
\newblock
\shownote{arXiv:2206.01825 [cs, econ, math, stat]}.


\bibitem[Chen(2007)]%
        {chen_chapter_2007}
\bibfield{author}{\bibinfo{person}{Xiaohong Chen}.} \bibinfo{year}{2007}\natexlab{}.
\newblock \showarticletitle{Chapter 76 Large Sample Sieve Estimation of Semi-Nonparametric Models}.
\newblock In \bibinfo{booktitle}{\emph{Handbook of Econometrics}}, \bibfield{editor}{\bibinfo{person}{James~J. Heckman} {and} \bibinfo{person}{Edward~E. Leamer}} (Eds.). Vol.~\bibinfo{volume}{6}. \bibinfo{publisher}{Elsevier}, \bibinfo{pages}{5549--5632}.
\newblock
\urldef\tempurl%
\url{https://doi.org/10.1016/S1573-4412(07)06076-X}
\showDOI{\tempurl}


\bibitem[Chen and Christensen(2015)]%
        {chen_optimal_2015}
\bibfield{author}{\bibinfo{person}{Xiaohong Chen} {and} \bibinfo{person}{Timothy~M. Christensen}.} \bibinfo{year}{2015}\natexlab{}.
\newblock \showarticletitle{Optimal uniform convergence rates and asymptotic normality for series estimators under weak dependence and weak conditions}.
\newblock \bibinfo{journal}{\emph{Journal of Econometrics}} \bibinfo{volume}{188}, \bibinfo{number}{2} (\bibinfo{date}{Oct.} \bibinfo{year}{2015}), \bibinfo{pages}{447--465}.
\newblock
\showISSN{0304-4076}
\urldef\tempurl%
\url{https://doi.org/10.1016/j.jeconom.2015.03.010}
\showDOI{\tempurl}


\bibitem[Chen and Shen(1998)]%
        {chen_sieve_1998}
\bibfield{author}{\bibinfo{person}{Xiaohong Chen} {and} \bibinfo{person}{Xiaotong Shen}.} \bibinfo{year}{1998}\natexlab{}.
\newblock \showarticletitle{Sieve {Extremum} {Estimates} for {Weakly} {Dependent} {Data}}.
\newblock \bibinfo{journal}{\emph{Econometrica}} \bibinfo{volume}{66}, \bibinfo{number}{2} (\bibinfo{year}{1998}), \bibinfo{pages}{289--314}.
\newblock
\showISSN{0012-9682}
\urldef\tempurl%
\url{https://doi.org/10.2307/2998559}
\showDOI{\tempurl}
\newblock
\shownote{Publisher: [Wiley, Econometric Society]}.


\bibitem[Chernozhukov et~al\mbox{.}(2018)]%
        {chernozhukov_doubledebiased_2018}
\bibfield{author}{\bibinfo{person}{Victor Chernozhukov}, \bibinfo{person}{Denis Chetverikov}, \bibinfo{person}{Mert Demirer}, \bibinfo{person}{Esther Duflo}, \bibinfo{person}{Christian Hansen}, \bibinfo{person}{Whitney Newey}, {and} \bibinfo{person}{James Robins}.} \bibinfo{year}{2018}\natexlab{}.
\newblock \showarticletitle{Double/debiased machine learning for treatment and structural parameters}.
\newblock \bibinfo{journal}{\emph{The Econometrics Journal}} \bibinfo{volume}{21}, \bibinfo{number}{1} (\bibinfo{date}{Feb.} \bibinfo{year}{2018}), \bibinfo{pages}{C1--C68}.
\newblock
\showISSN{1368-4221, 1368-423X}
\urldef\tempurl%
\url{https://doi.org/10.1111/ectj.12097}
\showDOI{\tempurl}


\bibitem[Chernozhukov et~al\mbox{.}(2022)]%
        {chernozhukov_automatic_2022}
\bibfield{author}{\bibinfo{person}{Victor Chernozhukov}, \bibinfo{person}{Whitney~K. Newey}, {and} \bibinfo{person}{Rahul Singh}.} \bibinfo{year}{2022}\natexlab{}.
\newblock \showarticletitle{Automatic {Debiased} {Machine} {Learning} of {Causal} and {Structural} {Effects}}.
\newblock \bibinfo{journal}{\emph{Econometrica}} \bibinfo{volume}{90}, \bibinfo{number}{3} (\bibinfo{year}{2022}), \bibinfo{pages}{967--1027}.
\newblock
\showISSN{1468-0262}
\urldef\tempurl%
\url{https://doi.org/10.3982/ECTA18515}
\showDOI{\tempurl}
\newblock
\shownote{\_eprint: https://onlinelibrary.wiley.com/doi/pdf/10.3982/ECTA18515}.


\bibitem[Corbae et~al\mbox{.}(2009)]%
        {corbae_introduction_2009}
\bibfield{author}{\bibinfo{person}{Dean Corbae}, \bibinfo{person}{Maxwell~B. Stinchcombe}, {and} \bibinfo{person}{Juraj Zeman}.} \bibinfo{year}{2009}\natexlab{}.
\newblock \bibinfo{booktitle}{\emph{An {Introduction} to {Mathematical} {Analysis} for {Economic} {Theory} and {Econometrics}}}.
\newblock \bibinfo{publisher}{Princeton University Press}.
\newblock


\bibitem[Davidson(2022)]%
        {davidson_stochastic_2022}
\bibfield{author}{\bibinfo{person}{James Davidson}.} \bibinfo{year}{2022}\natexlab{}.
\newblock \bibinfo{booktitle}{\emph{Stochastic {Limit} {Theory}: {An} {Introduction} for {Econometricians}} (\bibinfo{edition}{second edition, second edition} ed.)}.
\newblock \bibinfo{publisher}{Oxford University Press}, \bibinfo{address}{Oxford, New York}.
\newblock
\showISBNx{978-0-19-284450-7}


\bibitem[Dehling and Philipp(2002)]%
        {dehling_empirical_2002}
\bibfield{author}{\bibinfo{person}{Herold Dehling} {and} \bibinfo{person}{Walter Philipp}.} \bibinfo{year}{2002}\natexlab{}.
\newblock \showarticletitle{Empirical {Process} {Techniques} for {Dependent} {Data}}.
\newblock In \bibinfo{booktitle}{\emph{Empirical {Process} {Techniques} for {Dependent} {Data}}}, \bibfield{editor}{\bibinfo{person}{Herold Dehling}, \bibinfo{person}{Thomas Mikosch}, {and} \bibinfo{person}{Michael Sørensen}} (Eds.). \bibinfo{publisher}{Birkhäuser}, \bibinfo{address}{Boston, MA}, \bibinfo{pages}{3--113}.
\newblock
\showISBNx{978-1-4612-0099-4}
\urldef\tempurl%
\url{https://doi.org/10.1007/978-1-4612-0099-4_1}
\showDOI{\tempurl}


\bibitem[Doukhan(1994)]%
        {doukhan_mixing_1994}
\bibfield{author}{\bibinfo{person}{Paul Doukhan}.} \bibinfo{year}{1994}\natexlab{}.
\newblock \bibinfo{booktitle}{\emph{Mixing}}. \bibinfo{series}{Lecture {Notes} in {Statistics}}, Vol.~\bibinfo{volume}{85}.
\newblock \bibinfo{publisher}{Springer New York}, \bibinfo{address}{New York, NY}.
\newblock
\showISBNx{978-0-387-94214-8 978-1-4612-2642-0}
\urldef\tempurl%
\url{https://doi.org/10.1007/978-1-4612-2642-0}
\showDOI{\tempurl}


\bibitem[Eberlein(1984)]%
        {eberlein_weak_1984}
\bibfield{author}{\bibinfo{person}{Ernst Eberlein}.} \bibinfo{year}{1984}\natexlab{}.
\newblock \showarticletitle{Weak convergence of partial sums of absolutely regular sequences}.
\newblock \bibinfo{journal}{\emph{Statistics \& Probability Letters}} \bibinfo{volume}{2}, \bibinfo{number}{5} (\bibinfo{date}{Oct.} \bibinfo{year}{1984}), \bibinfo{pages}{291--293}.
\newblock
\showISSN{0167-7152}
\urldef\tempurl%
\url{https://doi.org/10.1016/0167-7152(84)90067-1}
\showDOI{\tempurl}


\bibitem[Engle et~al\mbox{.}(1986)]%
        {engle_semiparametric_1986}
\bibfield{author}{\bibinfo{person}{Robert~F. Engle}, \bibinfo{person}{C.~W.~J. Granger}, \bibinfo{person}{John Rice}, {and} \bibinfo{person}{Andrew Weiss}.} \bibinfo{year}{1986}\natexlab{}.
\newblock \showarticletitle{Semiparametric {Estimates} of the {Relation} {Between} {Weather} and {Electricity} {Sales}}.
\newblock \bibinfo{journal}{\emph{J. Amer. Statist. Assoc.}} \bibinfo{volume}{81}, \bibinfo{number}{394} (\bibinfo{year}{1986}), \bibinfo{pages}{310--320}.
\newblock
\showISSN{0162-1459}
\urldef\tempurl%
\url{https://doi.org/10.2307/2289218}
\showDOI{\tempurl}
\newblock
\shownote{Publisher: [American Statistical Association, Taylor \& Francis, Ltd.]}.


\bibitem[Fan and Huang(2005)]%
        {fan_profile_2005}
\bibfield{author}{\bibinfo{person}{Jianqing Fan} {and} \bibinfo{person}{Tao Huang}.} \bibinfo{year}{2005}\natexlab{}.
\newblock \showarticletitle{Profile likelihood inferences on semiparametric varying-coefficient partially linear models}.
\newblock \bibinfo{journal}{\emph{Bernoulli}} \bibinfo{volume}{11}, \bibinfo{number}{6} (\bibinfo{date}{Dec.} \bibinfo{year}{2005}), \bibinfo{pages}{1031--1057}.
\newblock
\showISSN{1350-7265}
\urldef\tempurl%
\url{https://doi.org/10.3150/bj/1137421639}
\showDOI{\tempurl}
\newblock
\shownote{Publisher: Bernoulli Society for Mathematical Statistics and Probability}.


\bibitem[Farrell et~al\mbox{.}(2021)]%
        {farrell_deep_2021}
\bibfield{author}{\bibinfo{person}{Max~H. Farrell}, \bibinfo{person}{Tengyuan Liang}, {and} \bibinfo{person}{Sanjog Misra}.} \bibinfo{year}{2021}\natexlab{}.
\newblock \showarticletitle{Deep {Neural} {Networks} for {Estimation} and {Inference}}.
\newblock \bibinfo{journal}{\emph{Econometrica}} \bibinfo{volume}{89}, \bibinfo{number}{1} (\bibinfo{year}{2021}), \bibinfo{pages}{181--213}.
\newblock
\showISSN{0012-9682}
\urldef\tempurl%
\url{https://doi.org/10.3982/ECTA16901}
\showDOI{\tempurl}
\newblock
\shownote{arXiv: 1809.09953}.


\bibitem[Gao(2007)]%
        {gao_nonlinear_2007}
\bibfield{author}{\bibinfo{person}{Jiti Gao}.} \bibinfo{year}{2007}\natexlab{}.
\newblock \bibinfo{booktitle}{\emph{Nonlinear {Time} {Series}: {Semiparametric} and {Nonparametric} {Methods}}}.
\newblock \bibinfo{publisher}{Chapman and Hall/CRC}, \bibinfo{address}{New York}.
\newblock
\showISBNx{978-0-429-19188-6}
\urldef\tempurl%
\url{https://doi.org/10.1201/9781420011210}
\showDOI{\tempurl}


\bibitem[Gao and Yee(2000)]%
        {gao_adaptive_2000}
\bibfield{author}{\bibinfo{person}{Jiti Gao} {and} \bibinfo{person}{Thomas Yee}.} \bibinfo{year}{2000}\natexlab{}.
\newblock \showarticletitle{Adaptive {Estimation} in {Partially} {Linear} {Autoregressive} {Models}}.
\newblock \bibinfo{journal}{\emph{The Canadian Journal of Statistics / La Revue Canadienne de Statistique}} \bibinfo{volume}{28}, \bibinfo{number}{3} (\bibinfo{year}{2000}), \bibinfo{pages}{571--586}.
\newblock
\showISSN{0319-5724}
\urldef\tempurl%
\url{https://doi.org/10.2307/3315966}
\showDOI{\tempurl}
\newblock
\shownote{Publisher: [Statistical Society of Canada, Wiley]}.


\bibitem[Geng et~al\mbox{.}(2020)]%
        {geng_estimation_2020}
\bibfield{author}{\bibinfo{person}{Xin Geng}, \bibinfo{person}{Carlos Martins-Filho}, {and} \bibinfo{person}{Feng Yao}.} \bibinfo{year}{2020}\natexlab{}.
\newblock \showarticletitle{Estimation of a partially linear additive model with generated covariates}.
\newblock \bibinfo{journal}{\emph{Journal of Statistical Planning and Inference}}  \bibinfo{volume}{208} (\bibinfo{date}{Sept.} \bibinfo{year}{2020}), \bibinfo{pages}{94--118}.
\newblock
\showISSN{03783758}
\urldef\tempurl%
\url{https://doi.org/10.1016/j.jspi.2020.02.002}
\showDOI{\tempurl}


\bibitem[Grenander(1981)]%
        {grenander_abstract_1981}
\bibfield{author}{\bibinfo{person}{Ulf Grenander}.} \bibinfo{year}{1981}\natexlab{}.
\newblock \bibinfo{booktitle}{\emph{Abstract {Inference}}}.
\newblock \bibinfo{publisher}{Wiley}.
\newblock
\showISBNx{978-0-471-08267-5}
\newblock
\shownote{Google-Books-ID: ng2oAAAAIAAJ}.


\bibitem[Härdle et~al\mbox{.}(2000)]%
        {hardle_partially_2000}
\bibfield{author}{\bibinfo{person}{Wolfgang Härdle}, \bibinfo{person}{Hua LIang}, {and} \bibinfo{person}{Jiti Gao}.} \bibinfo{year}{2000}\natexlab{}.
\newblock \bibinfo{booktitle}{\emph{Partially linear models}}.
\newblock \bibinfo{type}{{MPRA} {Paper}}. \bibinfo{institution}{University Library of Munich, Germany}.
\newblock
\urldef\tempurl%
\url{https://econpapers.repec.org/paper/pramprapa/39562.htm}
\showURL{%
\tempurl}


\bibitem[Kohler and Krzyżak(2017)]%
        {kohler_nonparametric_2017}
\bibfield{author}{\bibinfo{person}{Michael Kohler} {and} \bibinfo{person}{Adam Krzyżak}.} \bibinfo{year}{2017}\natexlab{}.
\newblock \showarticletitle{Nonparametric {Regression} {Based} on {Hierarchical} {Interaction} {Models}}.
\newblock \bibinfo{journal}{\emph{IEEE Transactions on Information Theory}} \bibinfo{volume}{63}, \bibinfo{number}{3} (\bibinfo{date}{March} \bibinfo{year}{2017}), \bibinfo{pages}{1620--1630}.
\newblock
\showISSN{1557-9654}
\urldef\tempurl%
\url{https://doi.org/10.1109/TIT.2016.2634401}
\showDOI{\tempurl}
\newblock
\shownote{Conference Name: IEEE Transactions on Information Theory}.


\bibitem[Kohler and Langer(2021)]%
        {kohler_rate_2021}
\bibfield{author}{\bibinfo{person}{Michael Kohler} {and} \bibinfo{person}{Sophie Langer}.} \bibinfo{year}{2021}\natexlab{}.
\newblock \showarticletitle{On the rate of convergence of fully connected deep neural network regression estimates}.
\newblock \bibinfo{journal}{\emph{The Annals of Statistics}} \bibinfo{volume}{49}, \bibinfo{number}{4} (\bibinfo{date}{Aug.} \bibinfo{year}{2021}), \bibinfo{pages}{2231--2249}.
\newblock
\showISSN{0090-5364, 2168-8966}
\urldef\tempurl%
\url{https://doi.org/10.1214/20-AOS2034}
\showDOI{\tempurl}
\newblock
\shownote{Publisher: Institute of Mathematical Statistics}.


\bibitem[Leadbetter et~al\mbox{.}(1983)]%
        {leadbetter_extremes_1983}
\bibfield{author}{\bibinfo{person}{M.~R. Leadbetter}, \bibinfo{person}{Georg Lindgren}, {and} \bibinfo{person}{Holger Rootzén}.} \bibinfo{year}{1983}\natexlab{}.
\newblock \bibinfo{booktitle}{\emph{Extremes and {Related} {Properties} of {Random} {Sequences} and {Processes}}}.
\newblock \bibinfo{publisher}{Springer}, \bibinfo{address}{New York, NY}.
\newblock
\showISBNx{978-1-4612-5451-5 978-1-4612-5449-2}
\urldef\tempurl%
\url{https://doi.org/10.1007/978-1-4612-5449-2}
\showDOI{\tempurl}


\bibitem[Li et~al\mbox{.}(2024)]%
        {li_simultaneous_2024}
\bibfield{author}{\bibinfo{person}{Jiaqi Li}, \bibinfo{person}{Likai Chen}, \bibinfo{person}{Kun~Ho Kim}, {and} \bibinfo{person}{Tianwei Zhou}.} \bibinfo{year}{2024}\natexlab{}.
\newblock \showarticletitle{Simultaneous inference of a partially linear model in time series}.
\newblock \bibinfo{journal}{\emph{Journal of Time Series Analysis}} (\bibinfo{year}{2024}).
\newblock
\showISSN{1467-9892}
\urldef\tempurl%
\url{https://doi.org/10.1111/jtsa.12781}
\showDOI{\tempurl}


\bibitem[Maurer(2016)]%
        {maurer_vector_contraction_2016}
\bibfield{author}{\bibinfo{person}{Andreas Maurer}.} \bibinfo{year}{2016}\natexlab{}.
\newblock \bibinfo{title}{A vector-contraction inequality for {Rademacher} complexities}.
\newblock
\newblock
\urldef\tempurl%
\url{http://arxiv.org/abs/1605.00251}
\showURL{%
\tempurl}
\newblock
\shownote{arXiv:1605.00251 [cs, stat]}.


\bibitem[Mokkadem(1988)]%
        {mokkadem_mixing_1988}
\bibfield{author}{\bibinfo{person}{Abdelkader Mokkadem}.} \bibinfo{year}{1988}\natexlab{}.
\newblock \showarticletitle{Mixing properties of {ARMA} processes}.
\newblock \bibinfo{journal}{\emph{Stochastic Processes and their Applications}} \bibinfo{volume}{29}, \bibinfo{number}{2} (\bibinfo{date}{Sept.} \bibinfo{year}{1988}), \bibinfo{pages}{309--315}.
\newblock
\showISSN{0304-4149}
\urldef\tempurl%
\url{https://doi.org/10.1016/0304-4149(88)90045-2}
\showDOI{\tempurl}


\bibitem[Newey(1990)]%
        {newey_efficient_1990}
\bibfield{author}{\bibinfo{person}{Whitney~K. Newey}.} \bibinfo{year}{1990}\natexlab{}.
\newblock \showarticletitle{Efficient {Instrumental} {Variables} {Estimation} of {Nonlinear} {Models}}.
\newblock \bibinfo{journal}{\emph{Econometrica}} \bibinfo{volume}{58}, \bibinfo{number}{4} (\bibinfo{year}{1990}), \bibinfo{pages}{809--837}.
\newblock
\showISSN{0012-9682}
\urldef\tempurl%
\url{https://doi.org/10.2307/2938351}
\showDOI{\tempurl}
\newblock
\shownote{Publisher: [Wiley, Econometric Society]}.


\bibitem[Robinson(1988)]%
        {robinson_1988}
\bibfield{author}{\bibinfo{person}{P.~M. Robinson}.} \bibinfo{year}{1988}\natexlab{}.
\newblock \showarticletitle{Root-{N}-{Consistent} {Semiparametric} {Regression}}.
\newblock \bibinfo{journal}{\emph{Econometrica}} \bibinfo{volume}{56}, \bibinfo{number}{4} (\bibinfo{year}{1988}), \bibinfo{pages}{931--954}.
\newblock
\showISSN{0012-9682}
\urldef\tempurl%
\url{https://doi.org/10.2307/1912705}
\showDOI{\tempurl}
\newblock
\shownote{Publisher: [Wiley, Econometric Society]}.


\bibitem[Schmidt-Hieber(2020)]%
        {schmidt_hieber_nonparametric_2020}
\bibfield{author}{\bibinfo{person}{Johannes Schmidt-Hieber}.} \bibinfo{year}{2020}\natexlab{}.
\newblock \showarticletitle{Nonparametric regression using deep neural networks with {ReLU} activation function}.
\newblock \bibinfo{journal}{\emph{The Annals of Statistics}} \bibinfo{volume}{48}, \bibinfo{number}{4} (\bibinfo{date}{Aug.} \bibinfo{year}{2020}).
\newblock
\showISSN{0090-5364}
\urldef\tempurl%
\url{https://doi.org/10.1214/19-AOS1875}
\showDOI{\tempurl}


\bibitem[Stone(1982)]%
        {stone_optimal_1982}
\bibfield{author}{\bibinfo{person}{Charles~J. Stone}.} \bibinfo{year}{1982}\natexlab{}.
\newblock \showarticletitle{Optimal {Global} {Rates} of {Convergence} for {Nonparametric} {Regression}}.
\newblock \bibinfo{journal}{\emph{The Annals of Statistics}} \bibinfo{volume}{10}, \bibinfo{number}{4} (\bibinfo{year}{1982}), \bibinfo{pages}{1040--1053}.
\newblock
\showISSN{0090-5364}
\urldef\tempurl%
\url{http://www.jstor.org/stable/2240707}
\showURL{%
\tempurl}
\newblock
\shownote{Publisher: Institute of Mathematical Statistics}.


\end{thebibliography}

                            \end{document}